\documentstyle[12pt,epsfig]{article}
\newcommand{\scaption}[1]{\caption{\protect{\footnotesize  #1}}}

\newcommand{\av}[1]{\mbox{$ \langle #1 \rangle $}}
\newcommand{\kjet}{\mbox{$k_{T\rm{jet}}$}}
\newcommand{\xjet}{\mbox{$x_{\rm{jet}}$}}
\newcommand{\ejet}{\mbox{$E_{\rm{jet}}$}}

\newcommand{\Q}{\mbox{$Q~$}}
\newcommand{\xb}{\mbox{$x~$}}  
\newcommand{\mz}{\mbox{$m_Z~$}}
\newcommand{\sigtot}{\mbox{$\sigma_{\rm tot}^{\gamma^* p}$}}
\newcommand{\sigt}{\mbox{$\sigma_T~$}}
\newcommand{\sigl}{\mbox{$\sigma_L~$}}

\newcommand{\Qsq}{\mbox{$Q^2~$}}
\newcommand{\et}{\mbox{$E_T~$}}
\newcommand{\kt}{\mbox{$k_T~$}}
\newcommand{\pt}{\mbox{$p_T~$}}
\newcommand{\ftwo}{\mbox{$F_2~$}}
\newcommand{\fl}{\mbox{$F_L~$}}

\newcommand{\as}{\mbox{$\alpha_s~$}}

\newcommand{\dif}{\mbox{\rm d}}

\newcommand{\GeV}{\mbox{\rm ~GeV~}}

\newcommand{\GeVsq}{\mbox{${\rm ~GeV}^2~$}}

\newcommand{\epem}{\mbox{$e^+e^-$}}

\begin{document}
%
%
%
\begin{titlepage}

%
\noindent
{\tt DESY 96-234    \hfill    ISSN 0418-9833} \\
{\tt MPI-PhE/96-23} \\
{\tt hep-ex/9611008} \\
{\tt November 1996}                  \\

\begin{center}


\vspace*{2cm}


{\bf  QCD AND THE STRUCTURE OF THE PROTON
\footnote{invited talk at the
XIX Workshop on High Energy Physics and Field Theory,
Protvino (Russia), June 1996} } \\
\vspace*{2.cm}
{\bf M. Kuhlen} \\ 
\vspace*{1.cm}
Max-Planck-Institut f\"ur Physik \\
Werner-Heisenberg-Institut \\
F\"ohringer Ring 6 \\
D-80805 M\"unchen  \\
Germany \\
E-mail: kuhlen@.desy.de
\bigskip
\bigskip
\\

\vspace*{2cm}

\end{center}

\begin{abstract}

\noindent

Measurements of structure functions and of 
the hadronic final state in deep inelastic
scattering at HERA are presented.
The results comprise the extraction of
parton densities, measurements of the strong
coupling, and the search for novel QCD 
effects in the new kinematic regime at HERA.

\vspace{1cm}

\end{abstract}
\end{titlepage}

\newpage

\section{Introduction}
%
%
\subsection{Overview}

Data from
the electron\footnote{here the generic
name electron includes the positron}-proton 
collider HERA at DESY in Hamburg provide new insights into
the structure of the proton. 
Electrons with $E_e=27 \GeV$ collide with protons of
$E_p=820 \GeV$, 
giving rise to a total 
$ep$ centre of mass (CM) energy of
$\sqrt{s} = \sqrt{4 E_e E_p} \approx 300 \GeV$. 
The high energy
allows to probe the proton
in hitherto unexplored kinematic regions.
Two complementary sources of information can be identified:
the measurement of the inclusive $ep \rightarrow e'H$
deep inelastic scattering (DIS) cross section, 
where 
$H$ stands for any hadronic system,
and measurements of the hadronic system $H$ itself. 
The cross section measurements are expressed in
structure functions, which in turn can be interpreted
in terms of parton (i.e. quark and gluon) densities in the 
proton. 
In simple words, how often the probe hits something
inside the proton determines its content.

The hadronic final state $H$ on the other hand
can be viewed as
the materialization of the quantum fluctuations inside the
proton, allowing in principle a more direct, 
but due to confinement also
a more complicated, access
to the dynamics inside the proton.
In addition, due to the large phase space,  
the hadronic final state serves as a QCD laboratory
with tunable initial kinematic conditions.

\subsection{Kinematics}

\begin{figure}[htb]
   \centering
   \begin{picture}(1,1) \put(0.,40.){QPM} \end{picture}
   \epsfig{file=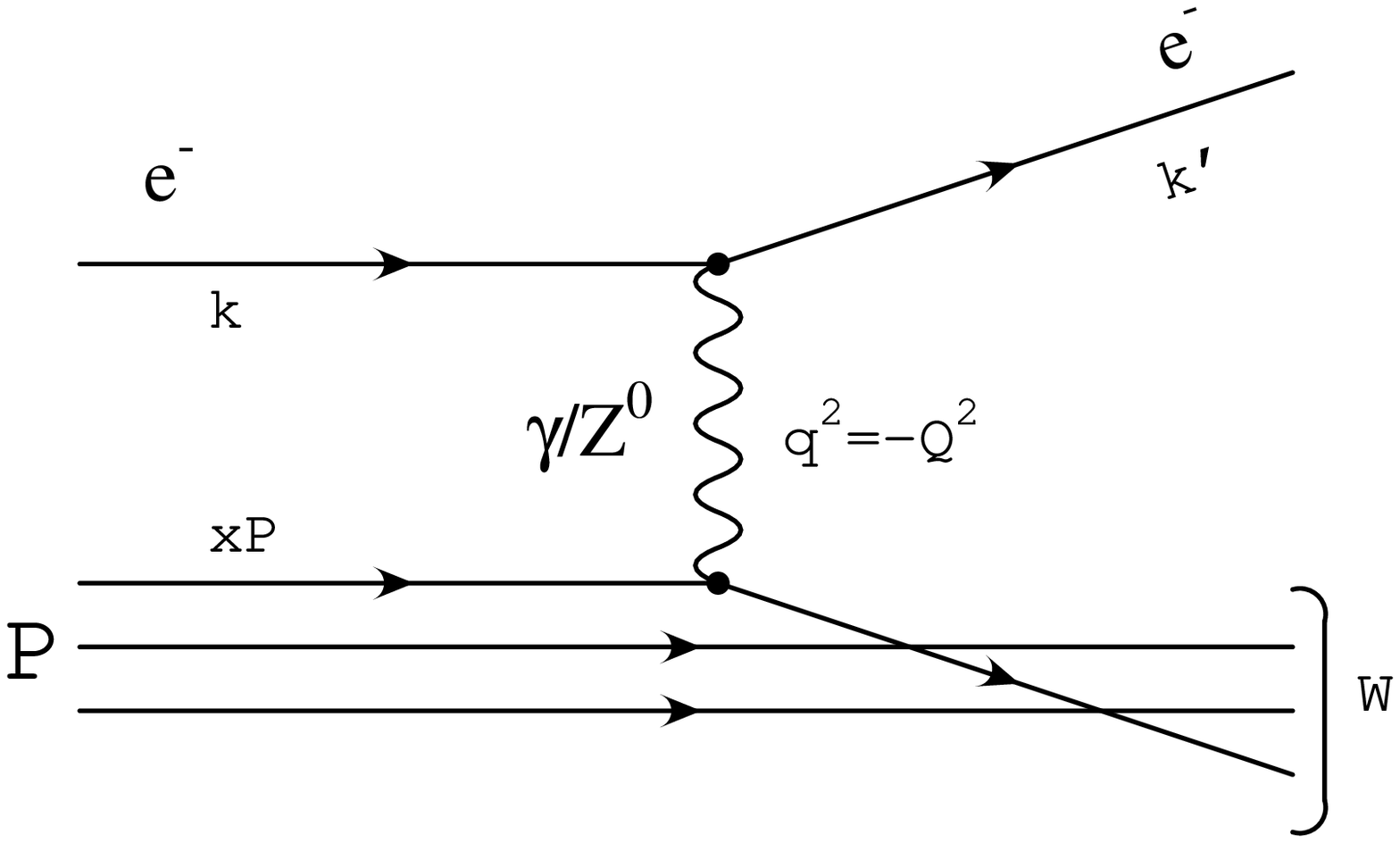,width=6cm,
    bbllx=50pt,bblly=483pt,bburx=522pt,bbury=771pt,clip=}
   \hspace{1cm}
   \epsfig{file=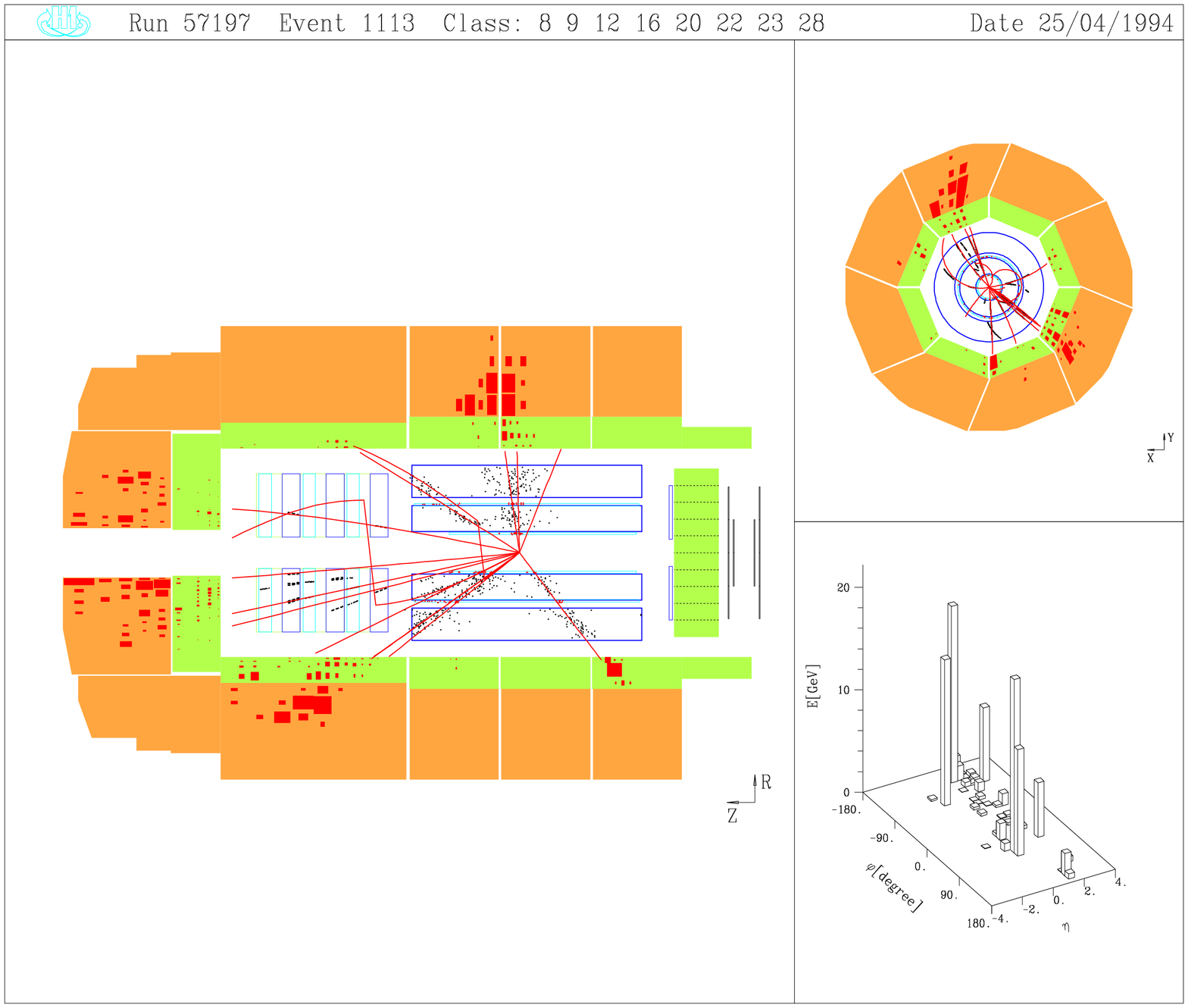,width=5cm,%
   bbllx=109pt,bblly=323pt,bburx=401pt,bbury=746pt,angle=90,clip=}
%
%
   \scaption{ {\bf a)}
              Basic diagram for DIS in 
              $O(\alpha_s^0)$ (quark parton model - QPM).
              {\bf b)}
              As an example, a DIS event  
              with the scattered electron and
              two well separated jets detected in the H1 detector is shown.
              The proton remnant leaves mostly 
              undetected in the $+z$ direction.}
%
   \label{kin} 
\end{figure}

The kinematics of the basic scattering process in Fig.~\ref{kin}
can be characterized by any set of two
Lorentz-invariants out of $Q^2,x,y,W$,
which are built from the 4-momentum
transfer $q=k-k'$ mediated by the virtual boson, and by the 
4-momentum $p$ of the incoming proton. These are:

   $\Qsq = -q^2$, which gives the resolving power
   of the probe with wavelength $\lambda =1/Q$ 
   (we set $\hbar = c = 1$);

   $x=\Qsq/(2pq)$, the Bjorken scaling variable, which can be
   interpreted as the momentum fraction of the proton which is
   carried by the struck quark;

   $y=\Qsq/(xs)$, the transferred energy fraction from the electron
   to the proton in the proton rest frame; and

   $W^2=\Qsq (1-x)/x+m_p^2 \approx sy-\Qsq$, 
   the invariant mass squared of the hadronic
   system $H$.

\begin{figure}[htb]
   \centering
   \epsfig{file=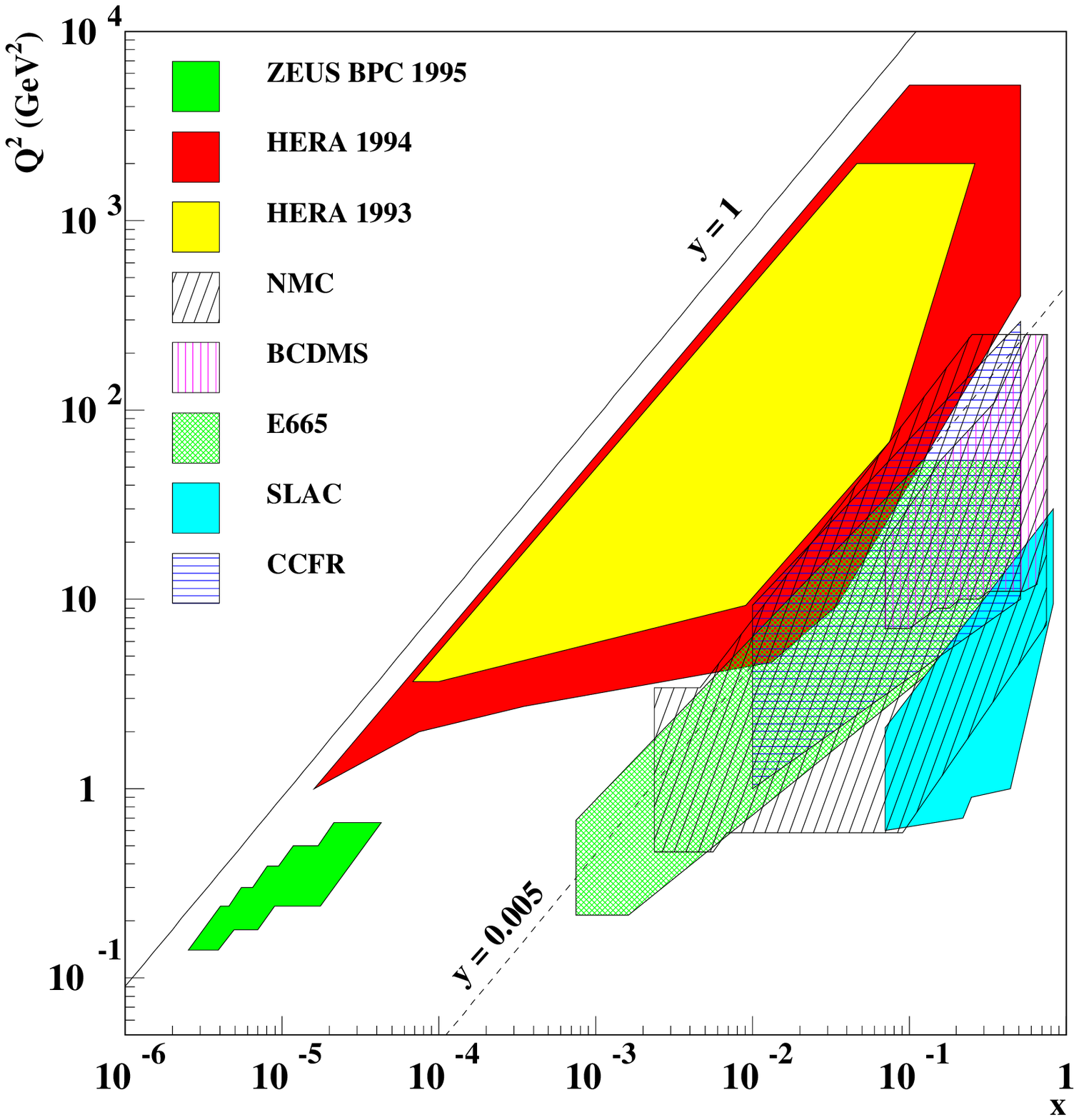,width=8cm,%
   bbllx=50pt,bblly=161pt,bburx=523pt,bbury=658}
   \scaption{Coverage in the kinematic plane $(x,\Qsq)$ 
             of various DIS experiments.}
   \label{hera_kine} 
\end{figure}

The large CM energy give access to 
kinematic regions both at very small $x$
and at large \Qsq (Fig. \ref{hera_kine}). The HERA data
cover roughly 
$\Qsq = 0.2 - 10^4 \GeVsq$,
$x=10^{-5} - 10^{-1}$ and
$W=40-300 \GeV$.

\subsection{Detectors}

Two large multi-purpose colliding beam detectors, H1 and ZEUS,
are installed around the collision regions (Fig.~\ref{kin}b). 
They consist mainly of inner tracking
devices, followed by calorimetry and outer muon detection systems.
They allow to measure both the scattered electron and the hadronic final state,
apart from the proton remnant, which largely escapes detection in the 
beam pipe. The scattered electron serves as a tag of DIS events, distinguishing
them from an otherwise overwhelming background from photoproduction
(quasi real photons with $\Qsq \approx 0$) and beam-gas reactions.
The kinematics can be determined either from the electron alone, or from
the measured hadronic system alone, or from a combination of both,
permitting important systematic cross checks.

\section{Structure functions}

The fundamental measurement in DIS concerns the cross section
for $ep\rightarrow e'H$ as a function of the kinematic variables.
The quark parton model (QPM) offers a physical picture:
the scattering takes place via a virtual photon which is
radiated off the scattering electron, and which couples
to a pointlike constituent inside the proton, that is 
a quark or antiquark.
The cross section is then proportional to the quark density inside
the proton. The general expression for the differential cross section
$ep\rightarrow e'H$ is
\begin{equation}
\frac{\dif^2\sigma}{\dif x \dif Q^2} =
\frac{4\pi \alpha^2}{xQ^4} \left[ \left(1-y+\frac{y^2}{2}\right) 
                             \cdot F_2(x,Q^2)
                          -  \frac{y^2}{2} \cdot F_L(x,Q^2) \right],
\label{eq:dsig}
\end{equation}  
where $Z$ exchange has been neglected (it is a 1\% correction
for $\Qsq\approx 1000 \GeVsq$).
The structure function \ftwo can be interpreted (in the ``DIS'' scheme) 
in terms 
of the quark and antiquark densities, $q_i$ and $\bar{q}_i$,
and their couplings to the photon, i.e. their charges $e_{q_i}$:
\begin{equation}
F_2(x,Q^2) = x \sum_{i} e_{q_i}^2 [q_i(x,Q^2) + \bar{q}_i(x,Q^2)],
\end{equation}
where the sum runs over all quark flavours. The longitudinal structure
function $F_L$ vanishes in zero'th order \as, and will be discussed later.

In the simple quark parton model the proton
consists just of 3 valence quarks. Their distribution functions in
fractional proton momentum \xb, $xq(x)$,  would peak at 
$\approx 1/3$ and tend
towards zero
for $x\rightarrow 0,1$.
In a static model of the proton, they would not depend on \Qsq.
It then follows that \ftwo should not depend on \Qsq, just on $x$ 
(Bjorken scaling).

When QCD is ``turned on'' the
quarks may radiate (and absorb)
gluons, which in turn may split into quark --
antiquark pairs or gluon pairs. 
More and more of these fluctuations can be resolved
with increasingly shorter wavelength of the photonic probe,
$\lambda = 1/Q$. 
With \Qsq increasing, we have a depletion of quarks
at large \xb, and a corresponding accumulation
at lower $x$.
In addition, ``sea quarks'' from $g\rightarrow q\bar{q}$
splittings populate small $x$. In fact, at small \xb it is
the gluon content with distribution function $g(x,Q^2)$
which governs the the proton and gives rise
to the DIS cross section via $q\bar{q}$ pairs.

The dynamic features of DIS have been calculated in 
various approximations of QCD.
One obtains equations for the \Qsq dependence of 
parton densities (and structure functions) like
\begin{equation}
  \frac{\dif q(x,Q^2)}{\dif \ln Q^2} = \frac{\alpha_s(Q^2)}{2\pi}
        \int_{x}^{1} \frac{dz}{z}
        \left[q(z,Q^2)P_{qq}(\frac{x}{z}) + 
              g(z,Q^2)P_{qg}(\frac{x}{z})\right], 
\label{eq:dglap}
\end{equation}
(and similarly for the gluon densities),
the famous
DGLAP (Dokshitzer-Gribov-Lipatov-Altarelli-Parisi)
equations \cite{th:dglap}.
They involve the calculable Altarelli-Parisi splitting functions
$P_{ij}(z)$, which 
give the probability of parton branchings
$q\rightarrow qg$, $g\rightarrow gg$ and  $g\rightarrow q\bar{q}$,
where the daughter parton $i$
carries a fraction $1-z$ of the mother's ($j$) momentum.
The coupled integro-differential equations for the quark and gluon
densities
can be solved, 
allowing to calculate them
for any value of $Q^2$, 
once they are known at a particular value $Q_0^2$.

For large $Q^2$ and for
small \xb where the proton is dominated by gluons
one obtains in the ``double leading log'' (DLL) approximation
($\ln\ln (Q^2/Q^2_0) \gg 1$, $\ln (1/x) \gg 1$) 
the formula \cite{th:dll}
\begin{equation}
  x g(x,Q^2) \approx x g(x,Q_0^2) \exp\sqrt{\frac{144}{25} 
 \ln \frac{\ln (Q^2/\Lambda^2)}{\ln (Q_0^2/\Lambda^2)}
 \ln (1/x) } .
\label{eq:dll}
\end{equation} 
At small \xb a fast rise of the gluon density is predicted.
That is, $xg$ increases faster than 
$(\ln \frac{1}{x})^\lambda$,
but slower than 
$(\frac{1}{x})^\lambda$
for any power $\lambda$,
the growth depending on the 
``evolution length'' from 
$Q_0^2$ to $Q^2$.

\subsection{The structure function \mbox{\boldmath \ftwo}}

\begin{figure}[h]
   \centering
   \epsfig{file=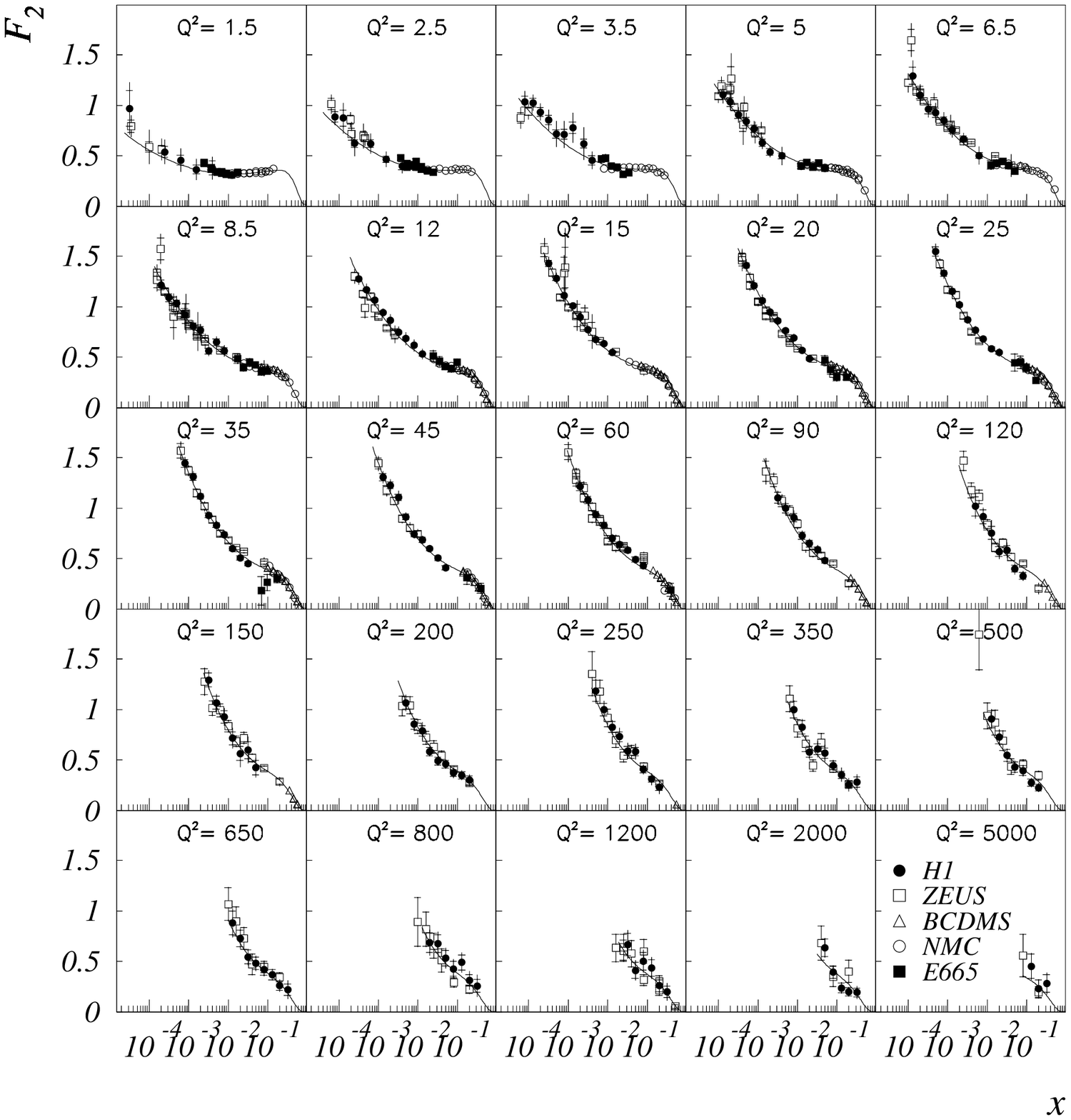,width=10cm,%
   bbllx=31pt,bblly=143pt,bburx=548pt,bbury=689pt}
   \scaption{The structure functions $F_2(x,Q^2)$ as a function
             of \xb for different \Qsq  bins. Shown are data
             from H1 \cite{h1:f2}, ZEUS \cite{z:f2},
             BCDMS \cite{o:bcdms}, NMC \cite{o:nmc} and
             E665 \cite{o:e665}.
             A NLO DGLAP QCD fit to the data with
             $Q^2>5\GeVsq$ is shown as full line.}
   \label{sfx} 
\end{figure}

ZEUS and H1 have measured \ftwo in a completely new kinematic domain
compared with fixed target experiments, most notably to much
smaller values of $x$.
The HERA data  \cite{h1:f2,z:f2} (Fig.~\ref{sfx})
exhibit a steep rise of \ftwo towards small \xb, which flattens at smaller
$Q^2$. They signal increasing parton densities at smaller 
$x$.

\begin{figure}[h]
   \centering
   \epsfig{file=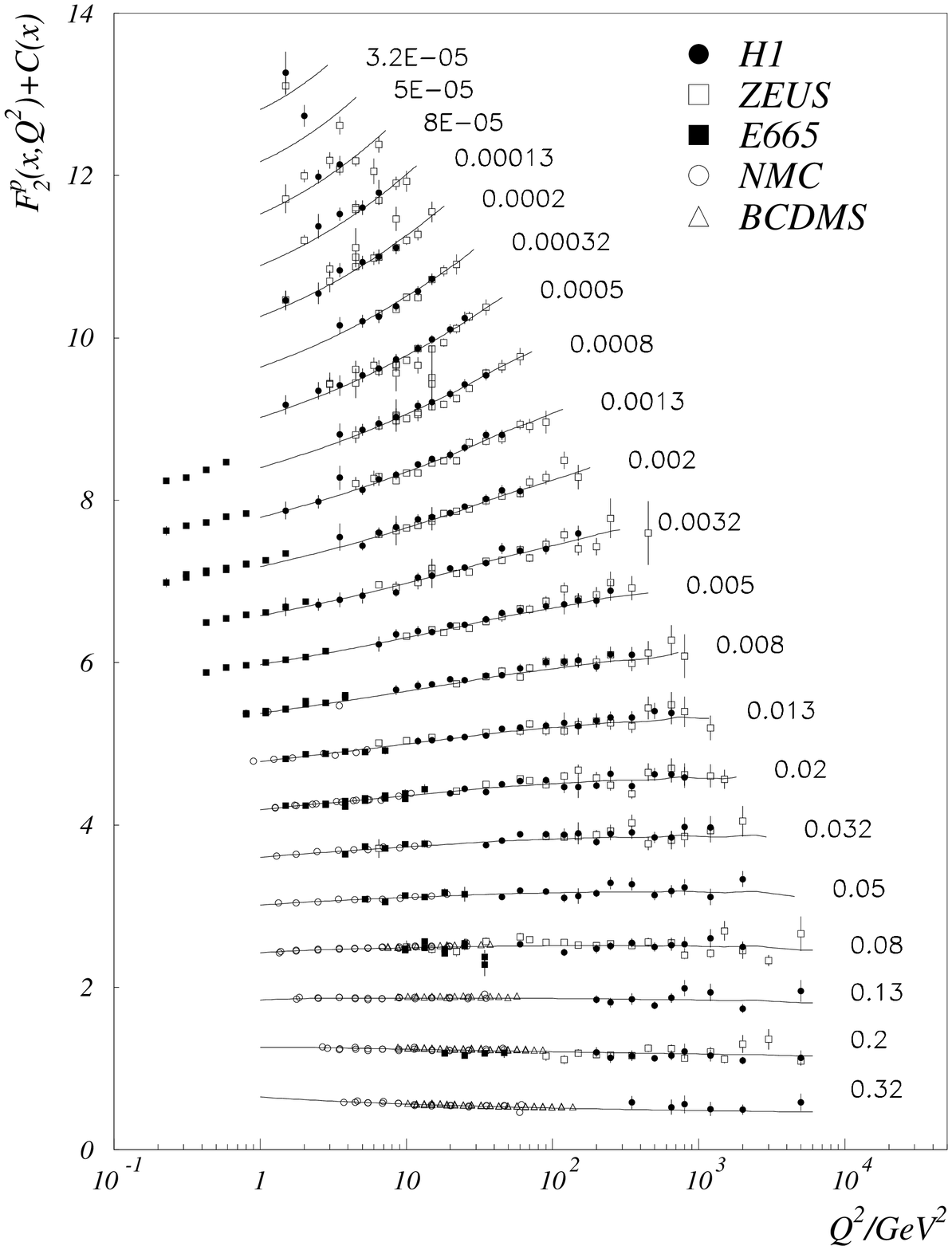,width=10cm,%
   bbllx=28pt,bblly=47pt,bburx=555pt,bbury=761pt}
   \scaption{The structure functions $F_2(x,Q^2)$ as a function
             of \Qsq for different \xb bins. 
             For better visibility, a function $C(x)=0.6(i-0.4)$,
             with $i$ the $x$ bin number ($i=1$ for $x=0.32$), is added
             to $F_2$ in the plot. Shown are the
             HERA and fixed target data, as well as a
             NLO QCD fit \cite{h1:f2}.}
   \label{sfq2} 
\end{figure}

In Fig.~\ref{sfq2} \ftwo is shown as a function of \Qsq for fixed \xb
values. At $x\approx 0.1$ scaling is observed -- \ftwo does not depend
on \Qsq. At $x>0.1$ \ftwo decreases with \Qsq due to parton splittings,
the products of which are found then at smaller \xb. Therefore
\ftwo increases with \Qsq for $x<0.1$ in accord with the above
qualitative discussion.

From a DGLAP evolution of pre-HERA data this 
sharp rise could not be predicted a priori,
because input distributions at small $x$ were not available.
It was known though that asymptotically for $\Qsq \rightarrow \infty$
the small $x$ behaviour is given by the DLL formula eq. \ref{eq:dll}.
The interesting question is whether the HERA data are consistent
with DGLAP evolution, or whether new effects have to be taken into account,
which one may expect at very small $x$.
The DGLAP evolution resums all 
leading terms $\sim(\as\ln (Q^2/Q_0^2))^n$,
but neglects terms $\sim(\as\ln (1/x))^n$ in the perturbation series.
The latter may become important at very small $x$ ($\as\ln (1/x) \gg 1$)
but moderate \Qsq, 
and are treated in
the BFKL (Balitsky-Fadin-Kuraev-Lipatov) \cite{th:bfkl} equation. 
The BFKL equation predicts an even more 
dramatic growth 
(the ``Lipatov growth'') of the gluon density than the DLL expression
at small $x$,
$F_2 \sim x g(x) ~ \sim x^{-\lambda}$, with $\lambda\approx0.5$ in 
leading order (LO).

It turns out, however, that the data can be fit perfectly well
with parton densities which obey the next-to-leading-order (NLO)
DGLAP evolution equations (see Fig.~\ref{sfx}) \cite{z:f2,h1:f2of94}. 
Standard QCD evolution
works over many orders of magnitude in both $x$ and $Q^2$!
H1 obtains values between 0.2 and 0.4 for the exponent in 
$F_2 \sim (1/x)^\lambda$, 
increasing with $Q^2$.
In fact, both the \xb and the \Qsq dependence of \ftwo can 
be attributed predominantly 
to the DLL formula,
which can be displayed nicely with
a suitable variable transformation 
(``double asymptotic scaling'' \cite{th:das}). 

It is remarkable that the \ftwo data can be
described by evolving flat or valence-like 
input quark and gluon distributions from 
a very low scale, $Q_0^2=0.35 \GeVsq$
up in \Qsq \cite{th:grv}. 
The steep rise with decreasing \xb is achieved by the 
long evolution length in $Q^2$ 
from the very low scale $Q_0^2=0.35 \GeVsq$ (see eq.~\ref{eq:dll}). 
The success of the GRV prediction came as a surprise
for many, as perturbative QCD should only be applicable for \Qsq
not too close to 
$\Lambda_{\rm QCD} \approx 0.2 \GeV$, because otherwise
$\as(Q^2) = 12 \pi / ((33-2n_f) \ln(Q^2/\Lambda_{\rm QCD}^2))$ 
diverges.
It is quite satisfying that the data can be described
with parton densities following standard QCD evolution.
However, the goal remains to calculate the measured
magnitude of the growth $(1/x)^\lambda$ from QCD,
rather than tuning it with the 
starting point $Q_0^2$ of the DLL evolution.

\subsection{QCD analysis of 
            \mbox{\boldmath \ftwo} --
           the gluon density and \mbox{\boldmath \as}}

From the DGLAP eq. \ref{eq:dglap} it is clear that 
the scaling violations of \ftwo depend on both \as and
the gluon density. In fact, for $x<0.01$ and in lowest order 
one can derive the approximate formula \cite{th:prytz}
\begin{equation}
  \frac{\dif F_2(x/2,\Qsq)}{\dif \ln \Qsq} \approx
       \frac{10}{27} \frac{\alpha_s (Q^2)}{\pi} xg(x,Q^2),
\end{equation}
because at small $x$ the proton is dominated by gluons, and the
scaling violations arise from quark pair creation from gluons.
The full NLO QCD analysis now employed at HERA is of course 
more involved.
Fig. \ref{qcdana}a shows
the gluon density $x\cdot g(x,Q^2)$
extracted from the NLO QCD fit to the \ftwo data.
It increases sharply towards small $x$,
the rise being more pronounced at larger $Q^2$.

Clearly, the density cannot increase forever; eventually
saturation effects will have to be taken into account
\cite{lowx:levin,lowx:mueller}. 
When gluons of transverse size $\sim 1/Q$
fill up the whole transverse
area offered by the proton, they will start to overlap and recombine.
This would be a novel
and very interesting situation indeed: high parton density,
but \Qsq large enough for $\as(Q^2)$ to be small! 
Given the size of the proton $\approx 1 {\rm fm}$, the critical
condition for saturation effects to turn on can be estimated 
\cite{lowx:mueller}
as 
$x_{\rm crit} g(x_{\rm crit},Q^2) \approx \frac{1 {\rm fm}^2}{1/Q^2}
                  \approx 25 \frac{Q^2}{\GeVsq}$. 
This value is by far
not reached by the measured gluon density (Fig.~\ref{qcdana}a).
It could be however that saturation does not set in uniformly over
the proton's transverse area, but starts locally in so-called
hot spots \cite{lowx:hotspots}. 
In this case $x_{\rm crit}$ would be larger.
The data however do not require any saturation correction.

From an analysis of the scaling violations 
$\dif \ftwo / \dif \ln \Qsq \sim \alpha_s$, or 
equivalently, of a QCD fit to the \ftwo data,
the strong coupling constant can be determined.
From an analysis of the 1993 data  
$\alpha_s(\mz) = 0.120 \pm 0.005 ({\rm exp.}) \pm 0.009({\rm theor.})$
was obtained \cite{th:ballforte}. This is being updated for
the higher statistics data now available \cite{th:as94}.


\subsection{The longitudinal structure function \fl}

The structure function \ftwo can also be expressed
in terms
of the cross sections
\sigt and \sigl for the absorption of transversely
and longitudinally polarized virtual
photons\footnote{we use the Hand convention \cite{th:hand}
for the definition of the virtual photon flux}
on protons:
$\sigtot = \sigl+\sigt$,
namely
\begin{equation}
\ftwo = \frac{Q^2(1-x)}{4\pi^2\alpha} 
\frac{Q^2}{Q^2+4 m_p^2 x^2} \cdot \sigtot
\approx \frac{\Qsq}{4\pi^2 \alpha}(\sigl+\sigt),
\end{equation}
where the small \xb approximation has been applied. 
Similarly,
\begin{equation}
F_L=\frac{Q^2 (1-x)}{4\pi^2\alpha} \cdot \sigl
\approx \frac{Q^2}{4\pi^2\alpha} \cdot \sigl.
\end{equation}
Longitudinal photons have helicity 0 and can exist only virtually.
In the QPM,
helicity
conservation at the electromagnetic vertex yields
the Callan-Gross relation, $F_L=0$, 
for scattering on quarks with spin $1/2$.
This does not hold when the quarks acquire transverse momenta
from QCD radiation. 
Instead, QCD yields \cite{th:am}
\begin{equation}
  F_L(x,Q^2)=\frac{\as}{4\pi} x^2 \int_{x}^{1}\frac{dz}{z^3}
      \left[\frac{16}{3}F_2(z,Q^2)
       +8\sum_i e_{q_i}^2(1-\frac{x}{z})\cdot g(z,Q^2)\right],
\end{equation}
exposing the dependence of \fl on the strong coupling and
the gluon density, which dominates over the first term at small $x$.
In fact 
$F_L(x,Q^2) = 0.7 \cdot [4\as/(3\pi)] \cdot xg(2.5x,Q^2)$ is not
a bad approximation for $x<10^{-3}$ \cite{th:fl}.

\begin{figure}[h]
   \centering
   \epsfig{file=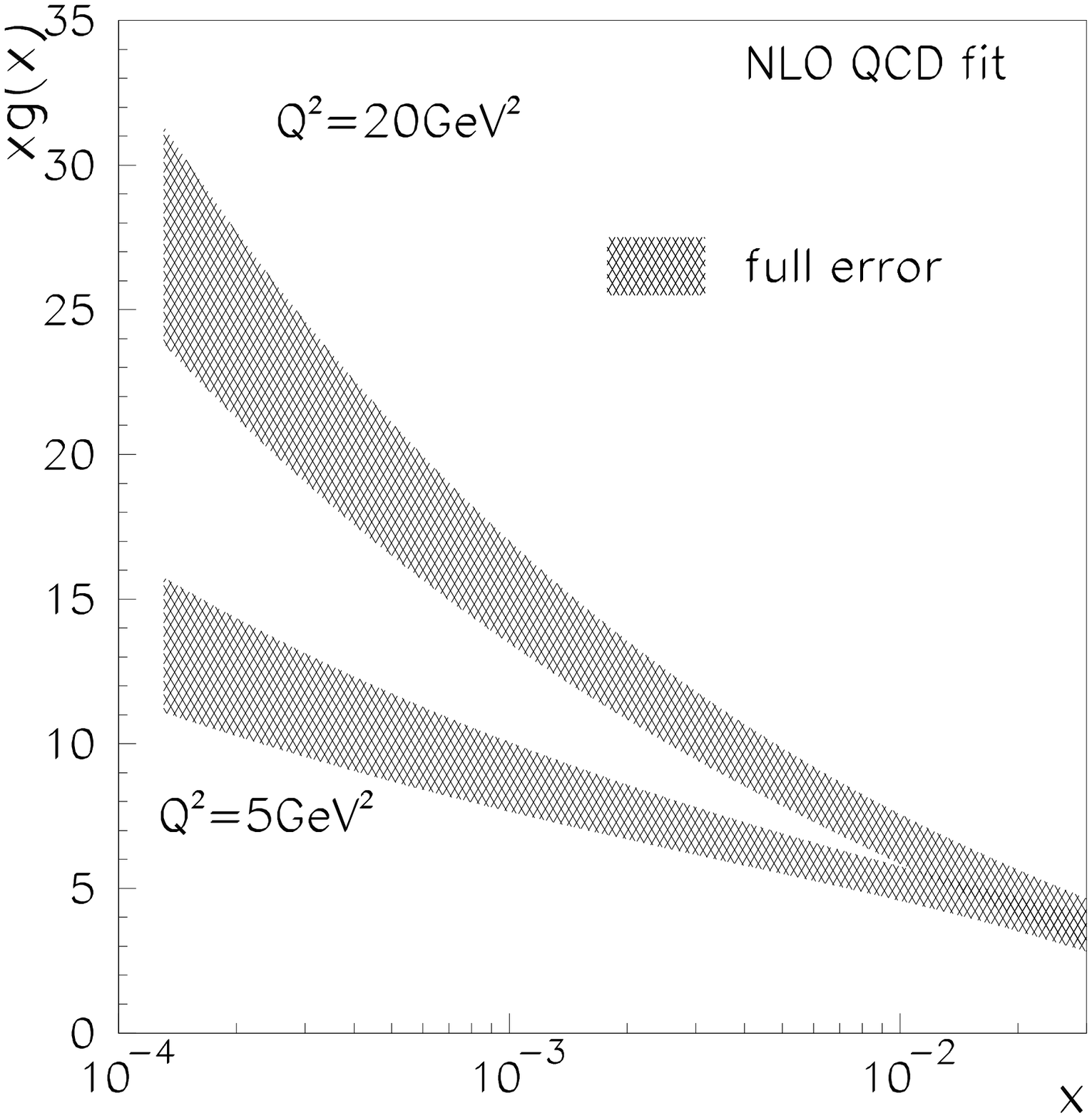,width=6cm,%
   bbllx=16pt,bblly=150pt,bburx=560pt,bbury=696}
   \epsfig{file=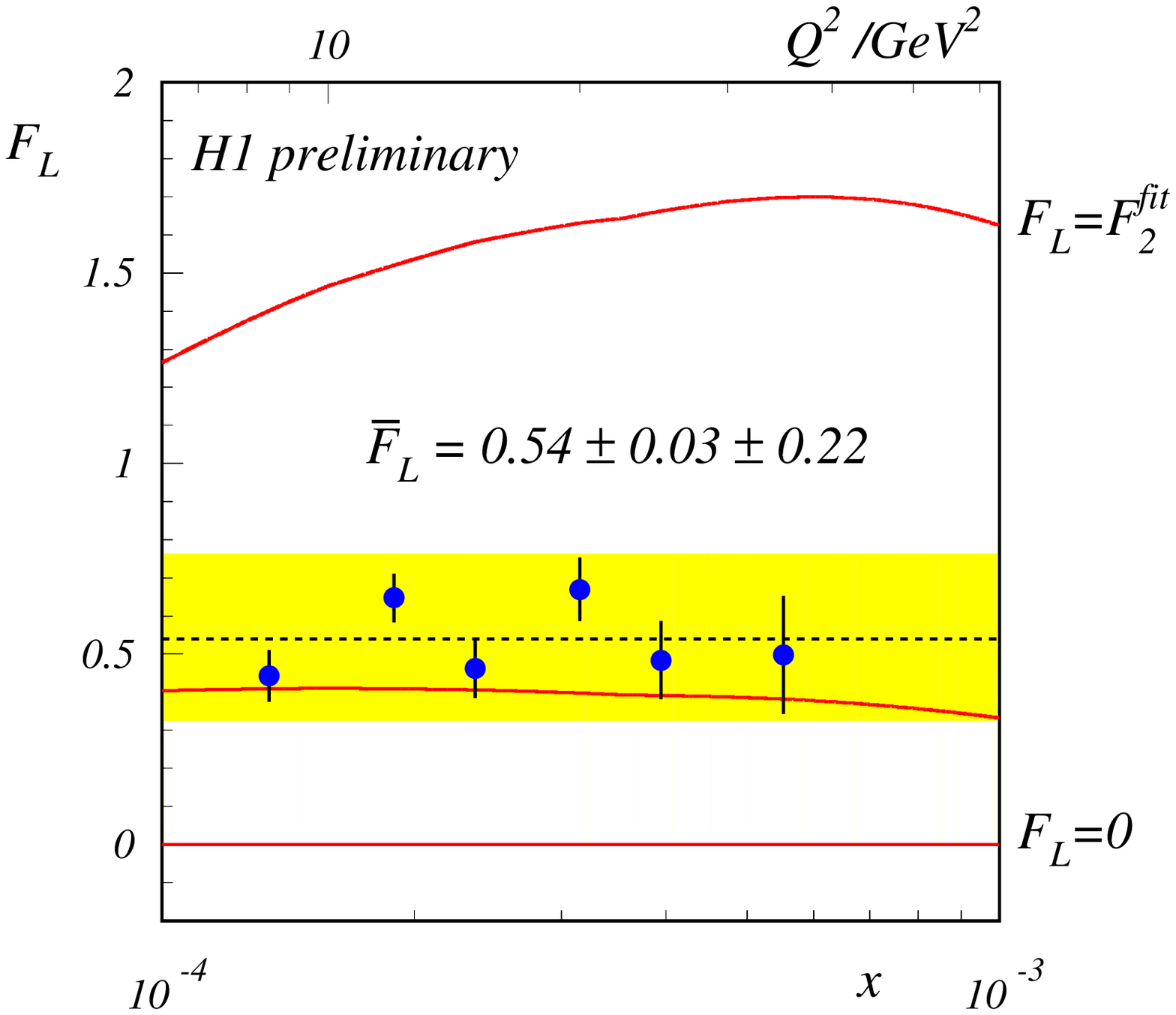,width=7cm,%
   bbllx=62pt,bblly=214pt,bburx=540pt,bbury=624}
   \scaption{{\bf a)} The gluon density $xg(x,Q^2)$ 
             as extracted from a QCD analysis
             of the H1 \ftwo data. 
             {\bf b)} The longitudinal structure functions $F_L$.
              The error bars represent statistical and systematic 
              point-to-point errors. The error band gives an 
              additional global
              systematic error. The QCD expectation based on the 
              GRV \cite{th:grv} parton densities is also shown
              (solid line, unlabelled).} 
   \label{qcdana} 
\end{figure}

The extraction of \ftwo from the cross section measurement 
(eq. \ref{eq:dsig})
so far had to make an assumption
for $F_L$, because there existed no data in the HERA regime. 
At large $y$, $y\approx 0.7$, this is a 10\% correction.
The argument can be turned around, and $F_L$ can be extracted
at large $y$ from a measurement of the cross section, assuming 
that \ftwo follows a QCD evolution and can be extrapolated from
measurements at smaller $y$. 
This procedure has been performed by H1 \cite{h1:fl} 
(Fig.~\ref{qcdana}b). 
The result $\fl=0.54\pm0.03\pm0.22$
excludes the extreme possibilities
\fl=\ftwo and \fl=0
and implies
$R \equiv \sigma_L/\sigma_T=F_L/(F_2-F_L) \approx 0.5$, 
since 
$F_2\approx 1.5$. This
is self consistent with the gluon density 
extracted from the H1 QCD fit.

\subsection{The very low \Qsq region}

\begin{figure}[h]
   \centering
   \epsfig{file=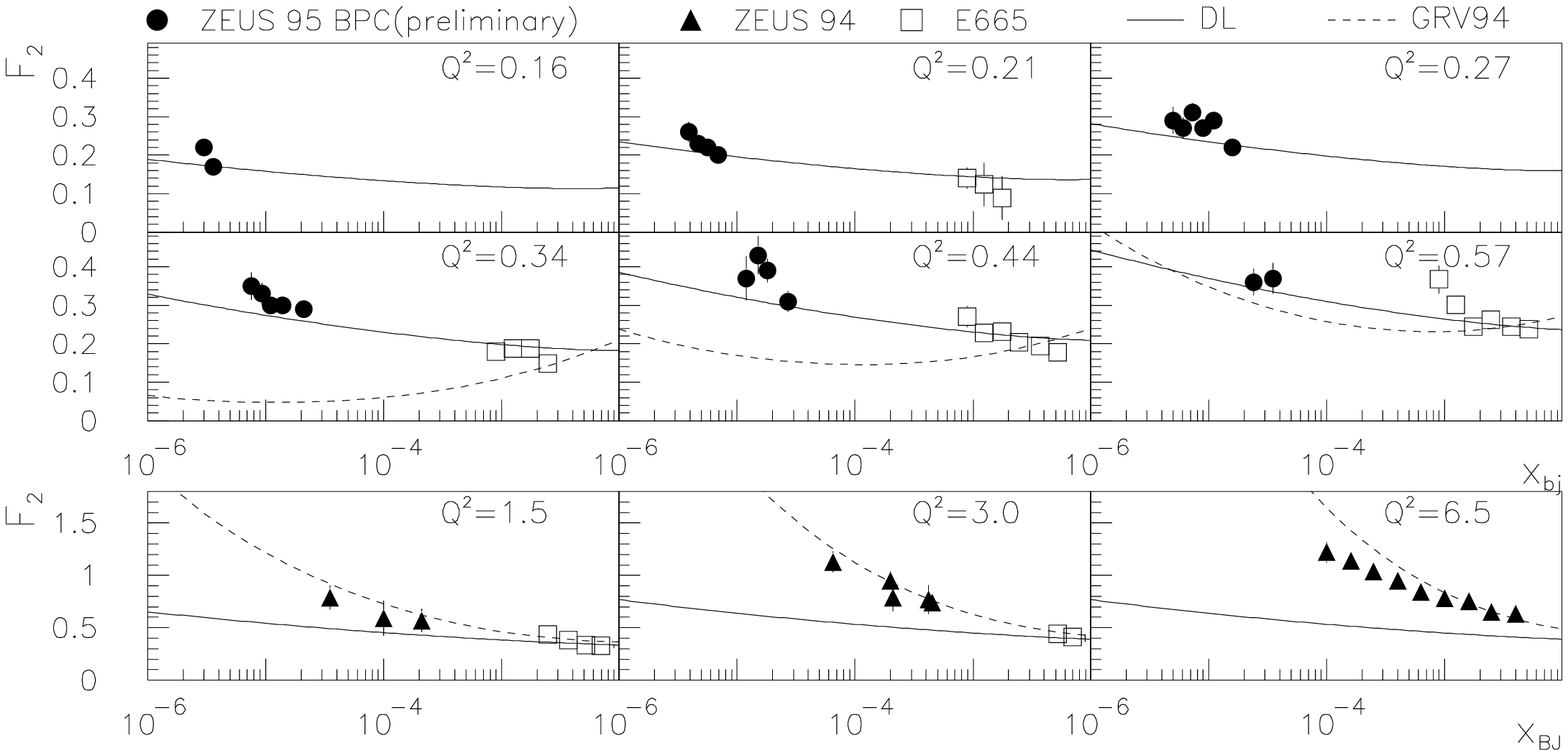,width=10cm,%
   bbllx=0pt,bblly=27pt,bburx=624pt,bbury=328}
   \scaption{The structure function $F_2$ measured at small $Q^2$ 
             by ZEUS \cite{z:f2bpc}.  
             Overlayed are the expectations from 
             the Donnachie-Landshoff \cite{th:dola2} Regge model (DL),
             and from the perturbatively evolved parton
             densities from Gl\"uck, Reya, Vogt \cite{th:grv} (GRV).}
   \label{zlowq2} 
\end{figure}

The ZEUS collaboration had for the 1994 run a new beam pipe calorimeter
installed to cover small electron scattering angles. 
They can now
access \Qsq values below 1~\GeVsq which had previously been the 
realm of fixed target experiments,
see Fig.~\ref{zlowq2} \cite{z:f2bpc}. 
Will perturbative QCD still hold at such a low scale? 
The perturbatively
evolved parton densities of GRV can describe the data down to 
$\Qsq \approx 1.5 \GeVsq$. 
Below 
$\Qsq\approx 0.4 \GeVsq$ the GRV
curves turn over and become valence like, by far failing 
to account for the data.
A non-perturbative model by Donnachie-Landshoff \cite{th:dola2}, 
which is based
on Regge phenomenology and reproduces the total photoproduction
cross section ($Q^2\approx0$) 
gives a good description of the data in this 
regime, but fails above $\Qsq\approx 2\GeVsq$.
The transition from perturbative to non-perturbative QCD
appears to be rather fast.

\section{The hadronic final state}

\begin{figure}[h]
   \centering
   \vspace{-0.2cm}
   \epsfig{file=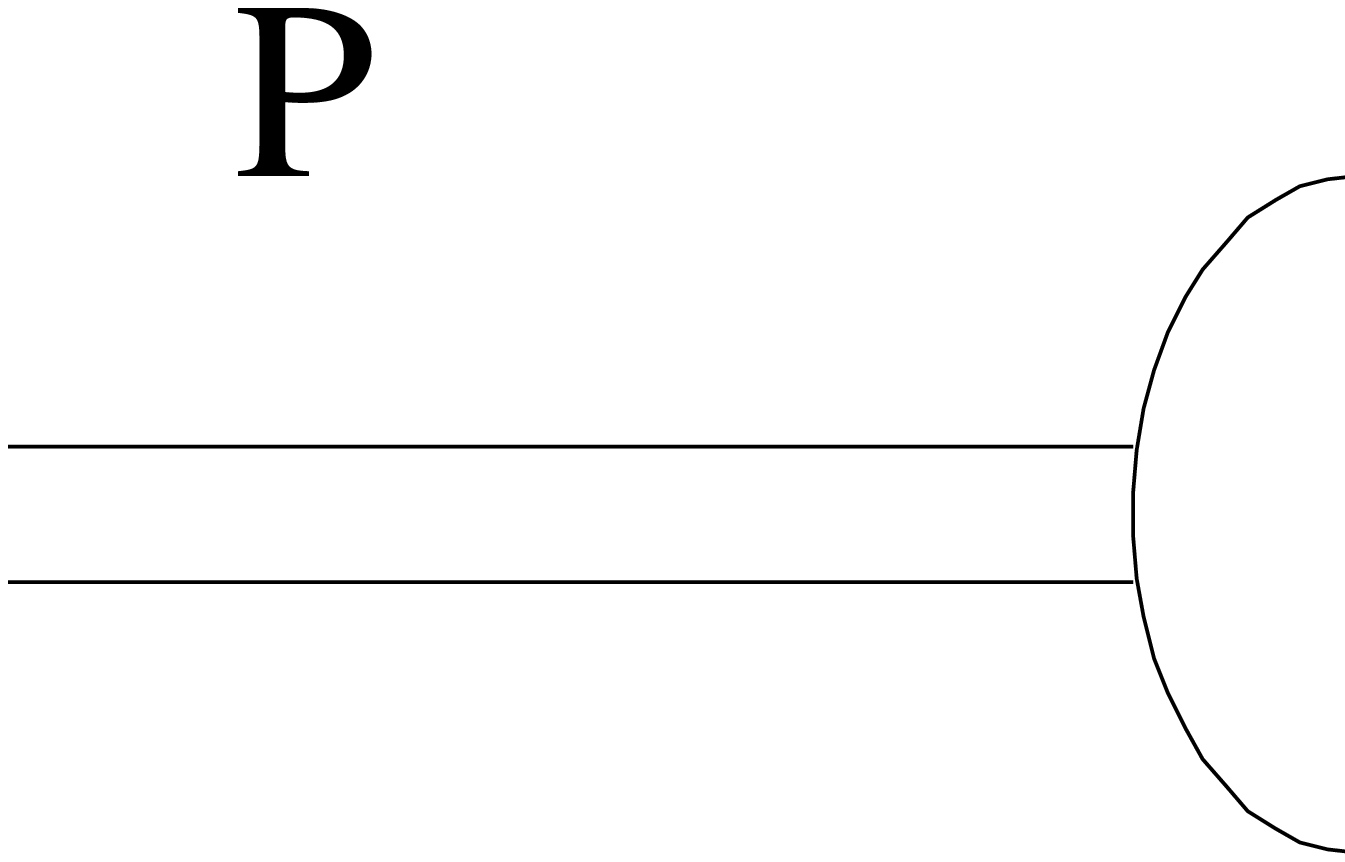,width=2cm}
   \hspace{3cm}
   \epsfig{file=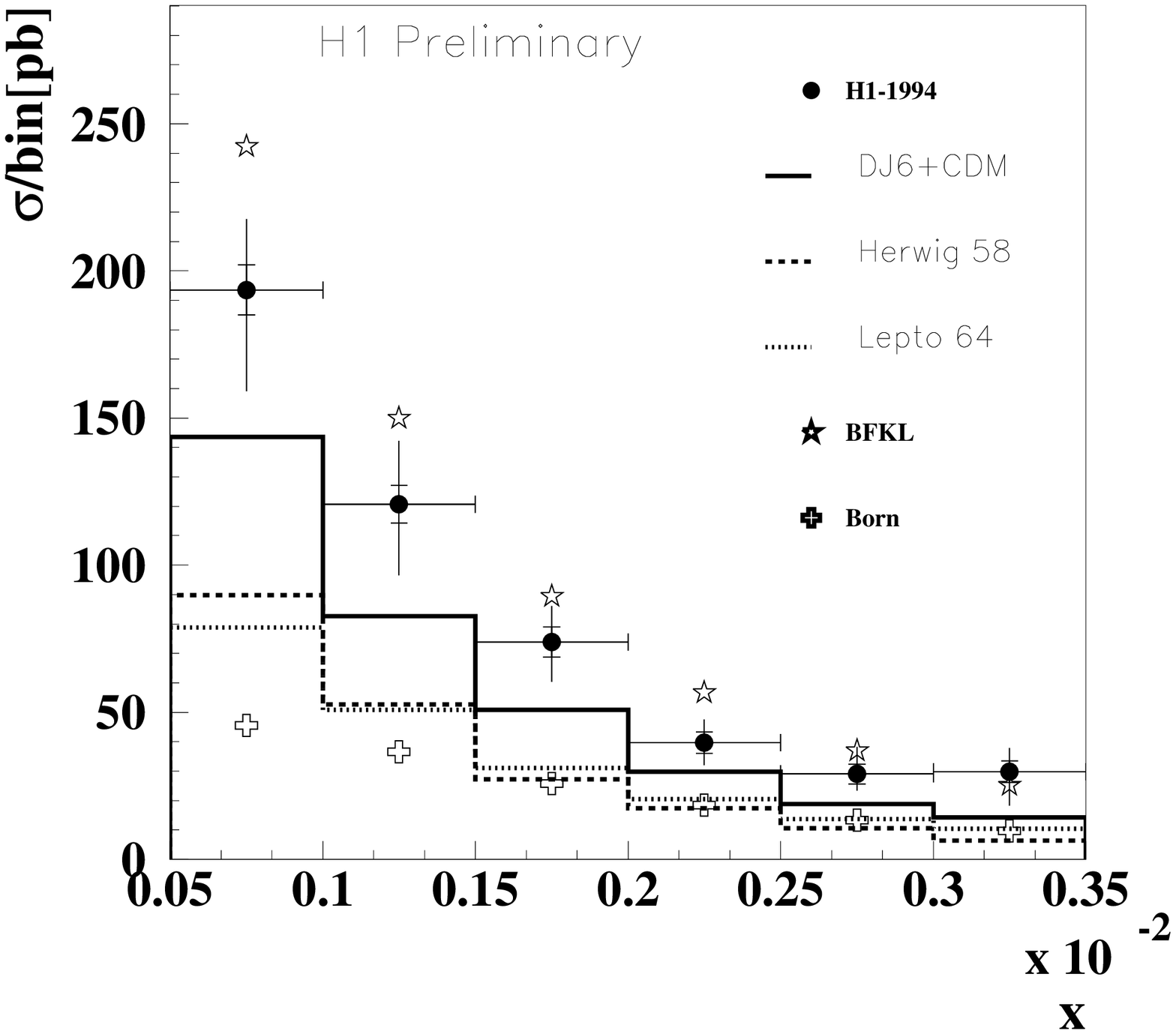,width=7cm,%
   bbllx=0pt,bblly=0pt,bburx=555pt,bbury=511,clip=}
   \scaption{ 
       {\bf a)} A ladder graph for parton evolution.
       The selection of forward jets in DIS events is illustrated.
       {\bf b)} The cross section for forward jets as a function
       of $x$. Overlayed are QCD model predictions for hadron jets,
       and BFKL and Born graph calculations for partons.
       The jets are 
       selected with $\xjet>0.035$, 
       $0.5 < \kjet^2/\Qsq <2$ and $\kjet>3.5 \GeV$.} 
   \label{ladder} 
\end{figure}

New insights into both perturbative QCD and
confinement can be gained by studying the hadronic final state,
thereby providing complementary information to the totally inclusive 
structure function 
measurements. 
For example,
footprints of BFKL dynamics are being searched for in especially devised
observables (\et flows, forward jets, high \pt particles).
\ftwo may be too inclusive to be sensitive to these effects, because
all possible paths in the parton evolution are integrated over.
Differences between DGLAP and BFKL evolution
can be expected for the hadronic final state,
because in DGLAP the phase space for parton radiation, basically given
by $W$,
is restricted
by strong ordering of the transverse momenta \kt of subsequently
emitted partons, which is not the case for BFKL (Fig.\ref{ladder}a).

There exists also a possibility to discover new QCD effects,
for example spectacular QCD instanton events.
At a 
more ``down to earth'' level, the large phase space for 
hard QCD radiation ($W$ up to 300 GeV) can be exploited for 
the extraction of the strong coupling constant
from jet rates and other final state obervables. 
For both cases, the searches for new effects and precision
measurements, a good understanding of the final state
dynamics in parton showers and hadronization is 
a prerequisite for meaningful results

Apart from the laboratory frame,
the hadronic centre of mass
system (CMS) and the Breit frame are used in the analyses.
The Breit frame
is defined by the condition that the virtual photon does not transfer
energy, only momentum. In the QPM picture the scattering quark would thus
just reverse the direction of its momentum of magnitude $Q/2$.
The
CMS is defined as the centre of mass system of the incoming
proton and the virtual boson, i.e. the CMS of the hadronic final
state with invariant mass $W$.
In both systems
the hemisphere defined by the virtual photon direction is referred
to as the current region, the other (containing the proton remnant)
as the target region.
The CMS current and target systems are back to back with momentum
$W/2$ each. 
Longitudinal and transverse quantities are calculated 
with respect to the
boson direction.

\subsection{QCD models}

The complexity of the hadronic final state
makes analytical QCD calculations often very difficult -- there
exist but a few.
Indispensable tools for the study of the hadronic final
state are therefore Monte Carlo models
which simulate the underlying dynamics.
They incorporate QCD evolution and parton radiation in
different approximations and utilize phenomenological models
for the non-perturbative hadronization phase.
The MEPS model (Matrix Element plus Parton Shower), 
an option of the LEPTO generator \cite{mc:lepto},
incorporates the QCD matrix elements up to first order, with additional
soft emissions generated by adding leading log parton showers.
In the colour dipole model (CDM, generator ARIADNE) 
\cite{mc:dipole,mc:ariadne}
radiation stems from
a chain of independently radiating dipoles formed by 
the colour charges. 
The HERWIG model \cite{mc:herwig}
is also based on leading log parton showers, with
additional matrix element corrections \cite{mc:seymour}.
The CDM description 
of gluon emission is similar to that of BFKL evolution 
to the extent that 
the gluons emitted by the dipoles
do not obey strong ordering in $k_T$~\cite{mc:bfklcdm}. 
In the MEPS and HERWIG models 
the partons emitted in the cascade
are strongly ordered in $k_T$, because they are based on 
leading log DGLAP parton showers. 

\subsection{Transverse energy flow}

\begin{figure}[h]
   \centering
   \epsfig{file=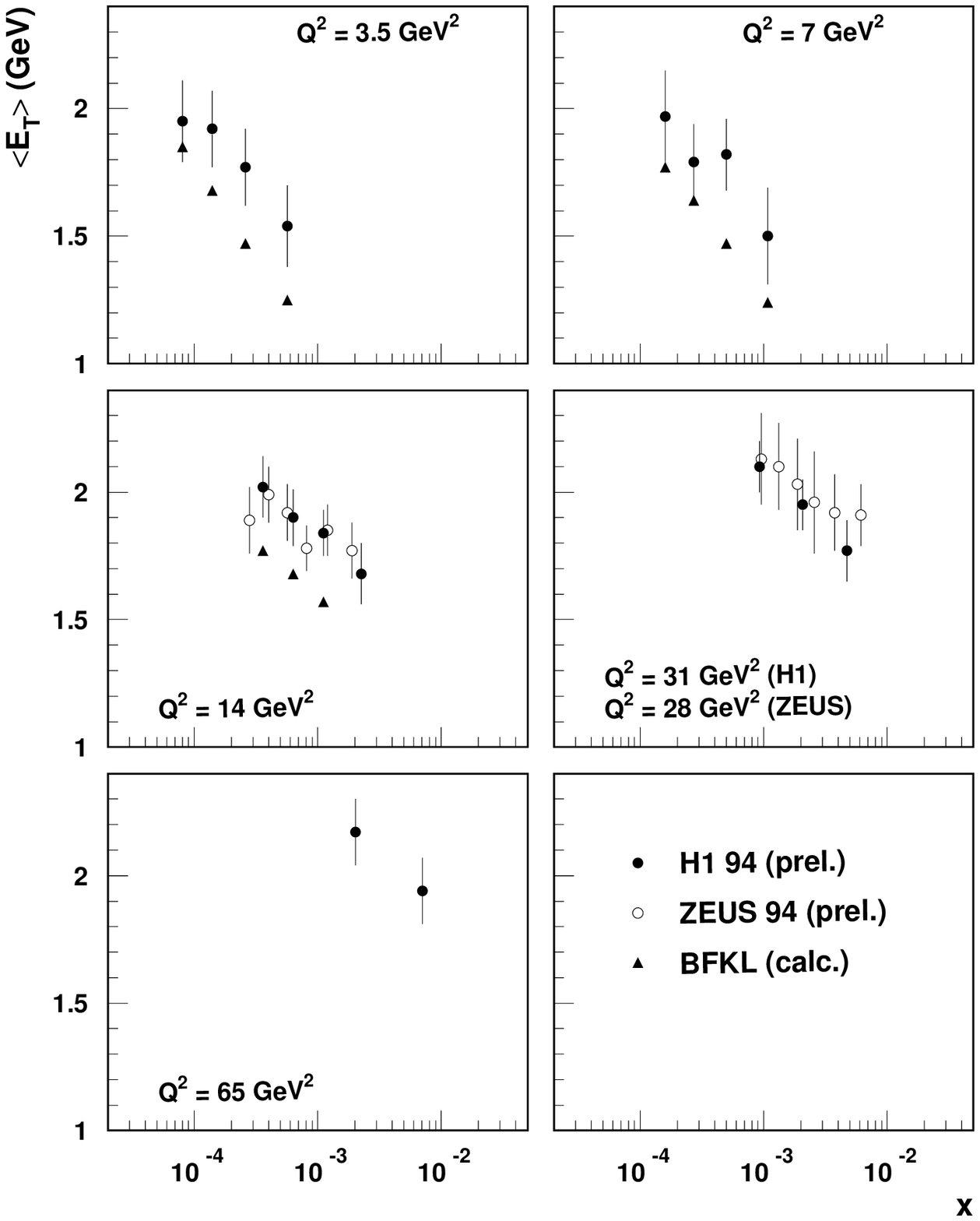,width=7cm,%
   bbllx=0pt,bblly=0pt,bburx=436pt,bbury=535}
   \epsfig{file=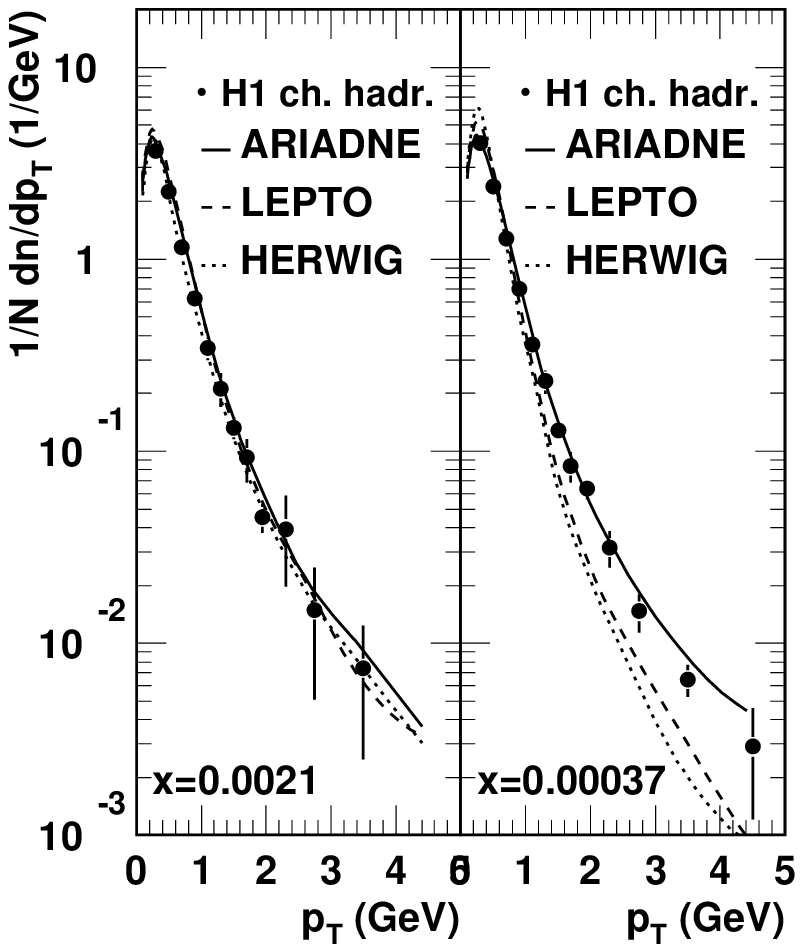,width=7cm,%
   bbllx=0pt,bblly=0pt,bburx=235pt,bbury=306,clip=}
   \scaption{
             {\bf a)} The average hadronic \et vs. $x$, measured in 
             $-0.5<\eta<0.5$ in the CMS
             from H1 \cite{h1:flow4} 
             and ZEUS \cite{z:flow} (both preliminary). 
             The BFKL
             calculation \cite{lowx:et} at the parton level 
             is also shown.
             {\bf b)} The charged particle \pt spectra measured in
              $0.5<\eta<1.5$ in the CMS by H1 for
             high and for low $x$ at
             $\Qsq \approx 14 \GeV$,  
             and the predictions from the 
             colour dipole model (ARIADNE), LEPTO and HERWIG.
             The plots are normalized to the total number of events $N$.} 
   \label{etpt} 
\end{figure}

It can be shown \cite{lowx:ordering} that 
DGLAP evolution corresponds to the evaluation of
ladder graphs (Fig.~\ref{ladder}) in which subsequently emitted gluons
are strongly ordered in transverse momenta \kt, 
$Q_0^2 \ll \kt_1^2 \ll ... \kt_j^2 \ll ... \Qsq$.
In the BFKL evolution on the other hand gluon emission
is not restricted by strong ordering, the emitted gluons
rather follow a kind of random walk in \kt space.
As a consequence, one expects more gluon radiation and hence more
transverse energy 
\et emitted in the rapidity region between
the remnant and the current systems
\cite{lowx:et}. 
The HERA data 
\cite{h1:flow4,z:flow} are compatible
with the predicted magnitude \cite{lowx:et} and the characteristic
increase towards small \xb from the BFKL calculation (Fig.~\ref{etpt}).
However, measured are hadrons, but the calculations do
not include hadronization!
It turns out that DGLAP models with large hadronization
corrections to the partonic \et are also able to describe the
data \cite{mc:sci,mc:carli}, precluding strong conclusions on
BFKL dynamics from the \et data.


\subsection{Inclusive transverse momentum spectra}

Another method to probe the low $x$ parton dynamics is to 
look for high \pt particles \cite{mk:method}.
They signal the presence of hard gluons, while hadronization
can only produce limited \pt.
Perturbative parton radiation can thus be disentangled from
hadronization effects, which pose a problem 
for the interpretation of the \et data.
H1 finds at low $x$ and at central rapidity
significantly harder \pt spectra than
expected from Monte Carlo calculations based upon
LO matrix element and leading log DGLAP parton
showers (Fig.~\ref{etpt}b). 
The colour dipole model, in which 
parton radiation is more abundant, gives a good description of the data.
QCD calculations and BFKL Monte Carlo predictions
for these spectra are eagerly awaited
to find out whether the enhanced parton radiation can indeed 
be attributed to BFKL evolution, or to something else, perhaps
even more interesting.  

\subsection{Forward jets}

A classic signature
for BFKL dynamics \cite{lowx:fwdjets,lowx:hotref} is the production of
``forward jets'' with $\xjet=\ejet / E_p$, the ratio of jet energy and
proton beam energy,
as large as possible, and with
transverse momentum \kjet ~close to \Q in order to reduce the phase
space for the \kt ordered DGLAP evolution (s. Fig.~\ref{ladder}).
An enhanced rate of events with such jets is thus expected in the BFKL
scheme \cite{lowx:fwdjets}.
The experimental difficulty is to detect these
``forward'' jets which are close to the beam hole in the proton
direction. 
The rate of forward jets measured by H1 \cite{h1:fwdjet}
(Fig.~\ref{ladder}b)
increases with falling $x$.
This is expected from BFKL
calculations, in contrast
to calculations without the BFKL ladder 
\cite{lowx:fwdcalc}. 
The behaviour of the data is better represented by the CDM 
than by the DGLAP representative, the MEPS model.
However,
the cross sections are calculated
\cite{lowx:fwdcalc} for partons, while
experiments measure
hadron jets. 
This gap has to be bridged from both sides to 
allow a strictly valid comparison. 

\subsection{Instantons}

\begin{figure}[h]
   \centering
\mbox{
   \epsfig{file=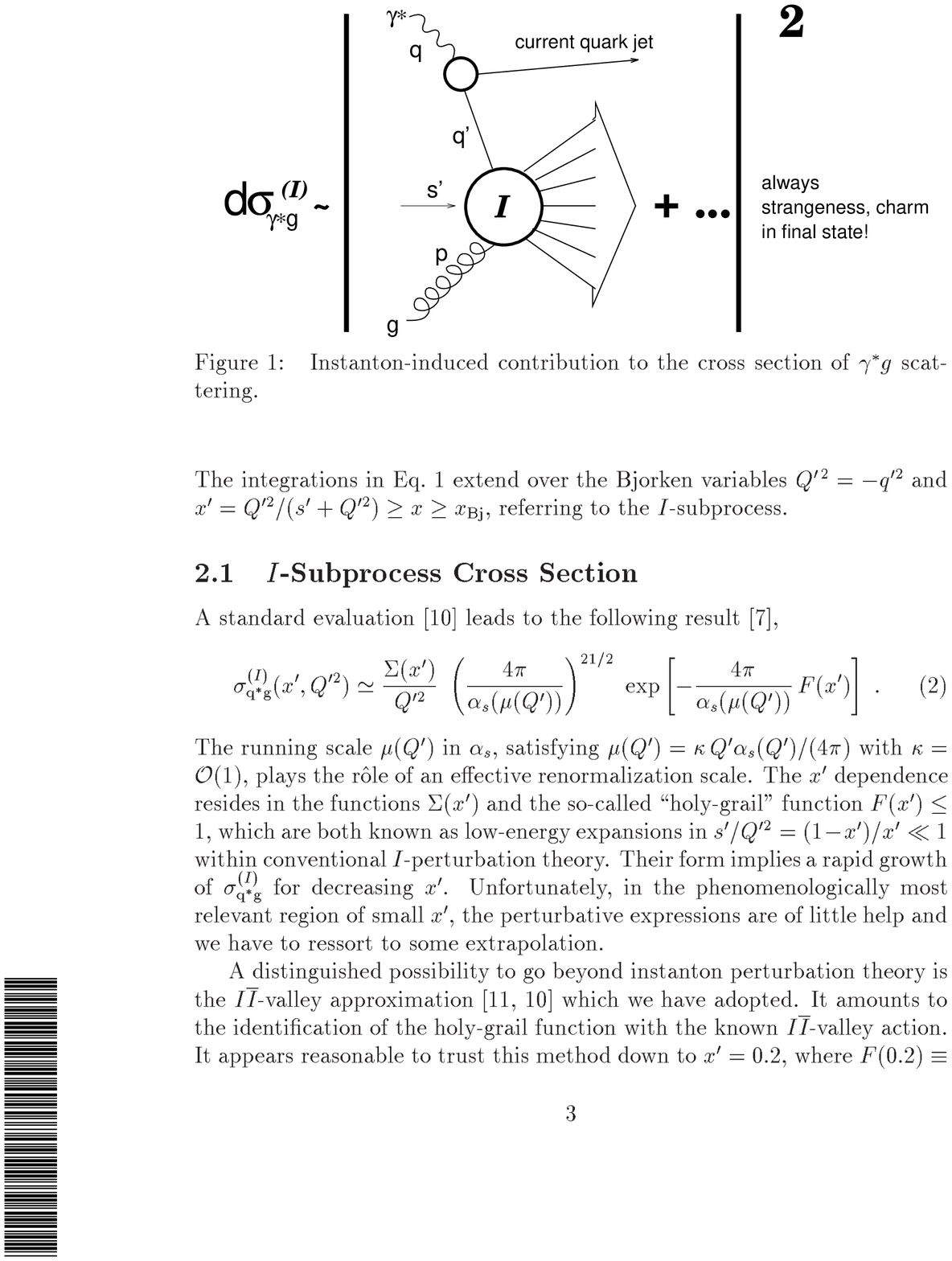,width=3.0cm,%
   bbllx=197pt,bblly=491pt,bburx=346pt,bbury=665,clip=}
}
   \epsfig{file=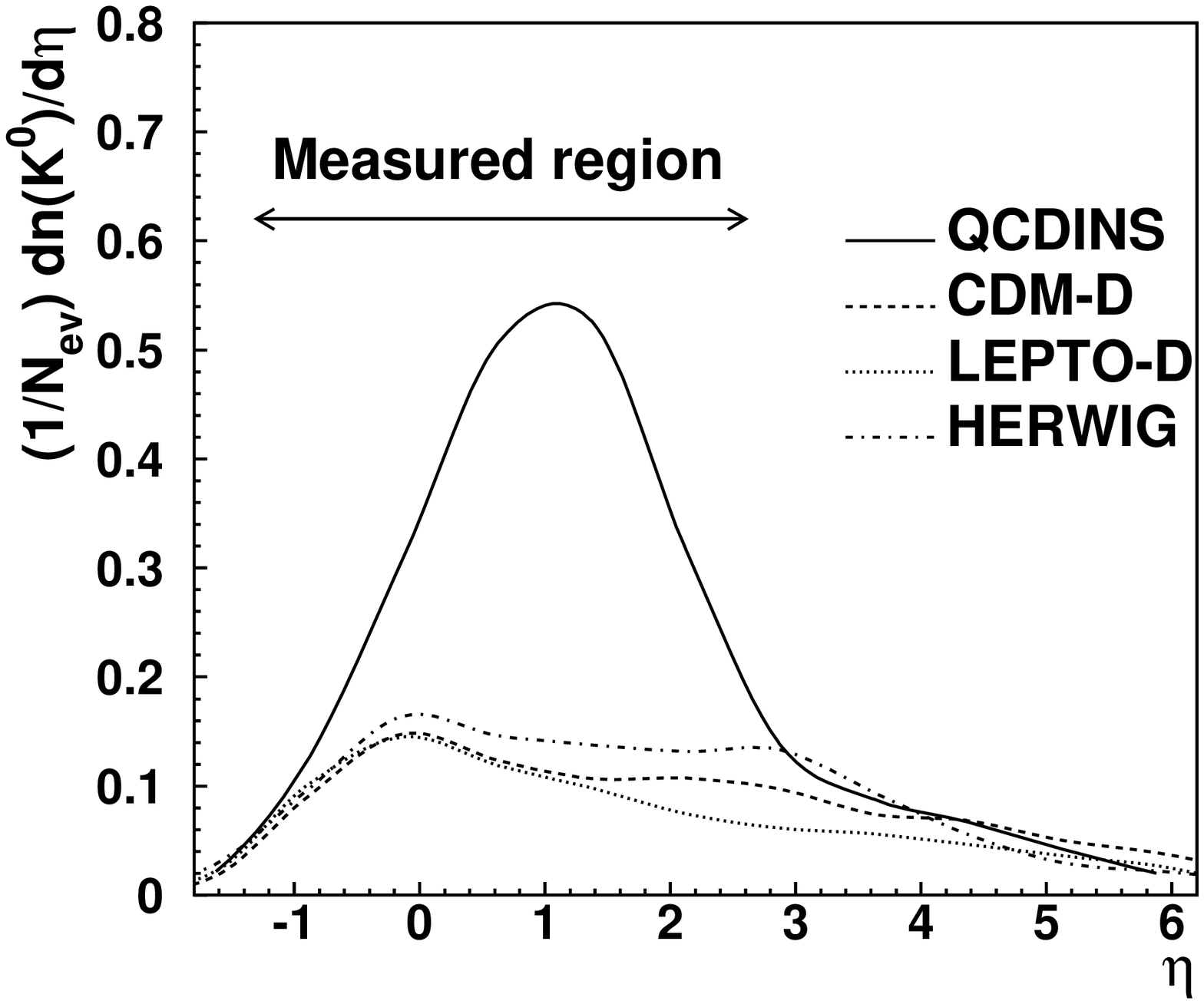,width=5.0cm,%
   bbllx=31pt,bblly=182pt,bburx=560pt,bbury=640}
   \epsfig{file=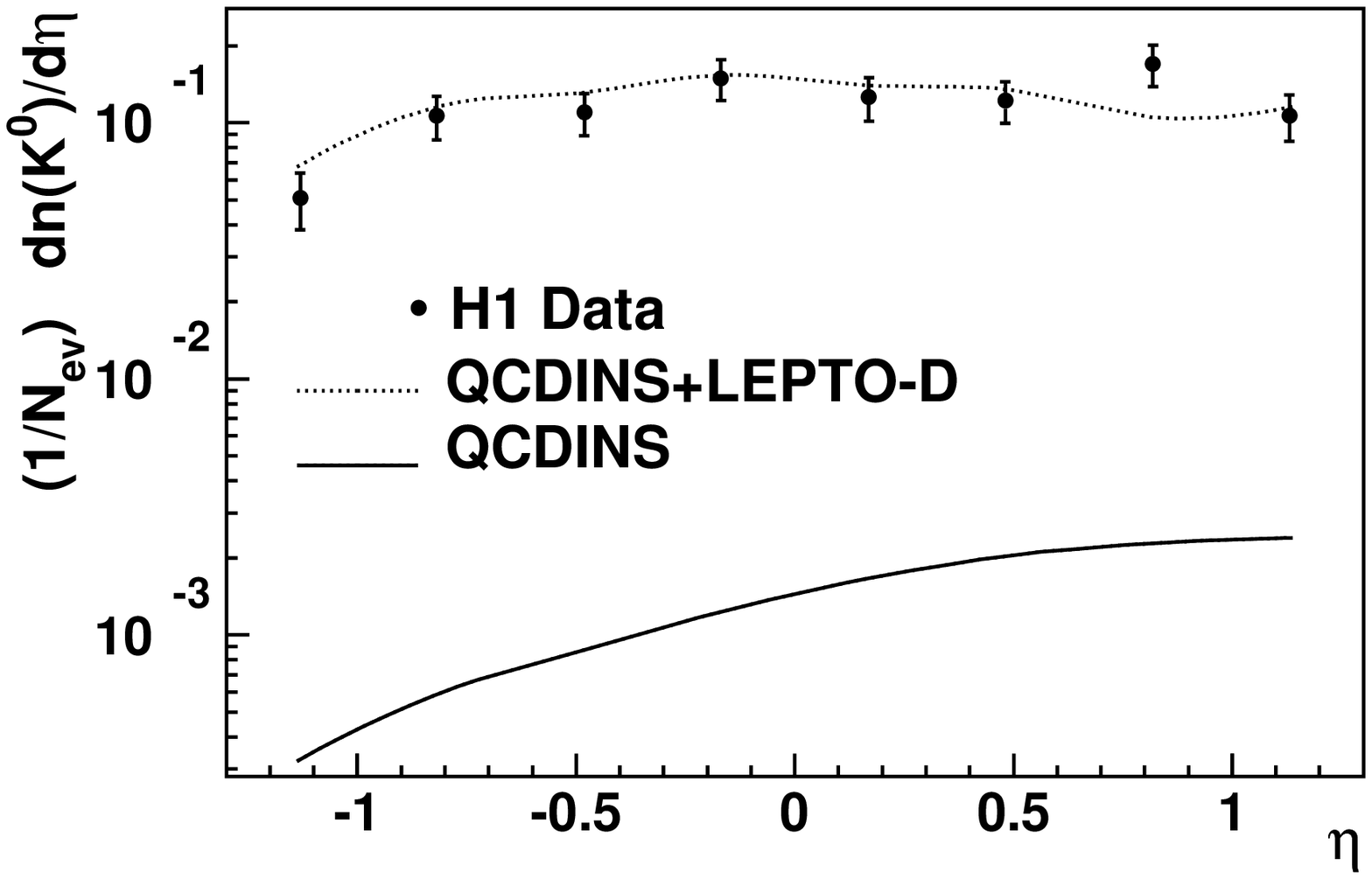,width=6.0cm,%
   bbllx=31pt,bblly=260pt,bburx=523pt,bbury=585}
   \scaption{ 
              {\bf a)} Instanton induced process, and the $K^0$ yield
              as a function of laboratory pseudorapidity $\eta$.
              The proton direction is to the right.
              Shown in {\bf b)} are the predictions from standard
              QCD models, and from an instanton model (QCDINS). 
              In {\bf c)} the H1 data \cite{h1:k0} are overlayed
              with the prediction of a standard QCD model plus 
              a 0.6\% instanton contribution, 
              and with the maximally allowed fraction of instanton events
              (QCDINS).} 
   \label{inst} 
\end{figure}

The
standard model contains processes which cannot be 
described by perturbation theory, and which violate
classical conservation laws like baryon and lepton number
in the case of the electroweak sector and chirality for
the strong interaction \cite{th:thooft}. 
Such anomalous processes are induced by instantons \cite{th:belavin}.
At HERA, QCD instantons may lead to observable effects
in the hadronic final state in DIS \cite{th:ringwald,th:schrempp}.
The instanton should decay isotropically into a high
multiplicity state, consisting of gluons
and all quark flavours (in each event!)  
which are kinematically allowed.
Due to the isotropic decay, one expects a densely populated
region in rapidity, other than the current jet, which is
isotropic in azimuth. The presence of strangeness and charm
provides an additional signature (Fig.~\ref{inst}a).
H1 does not see an excess of $K^0$ production
over the prediction from standard QCD models (Fig.~\ref{inst}) 
\cite{h1:k0}.
At most a few percent admixture of instanton events is allowed
to normal DIS, corresponding to an upper limit for the instanton
cross section for $x>0.001$
of 0.95 nb.
In the future,
more elaborate search strategies and larger luminosities offer
the chance for a fundamental discovery at HERA \cite{th:wsinst}!

\subsection{The running coupling constant 
            \mbox{\boldmath \as} from jet rates}

\begin{figure}[h]
   \centering
   \begin{picture}(1,1) \put(10.,30.){BGF} \end{picture}
   \begin{picture}(1,1) \put(140.,30.){QCDC} \end{picture}
\mbox{
   \epsfig{file=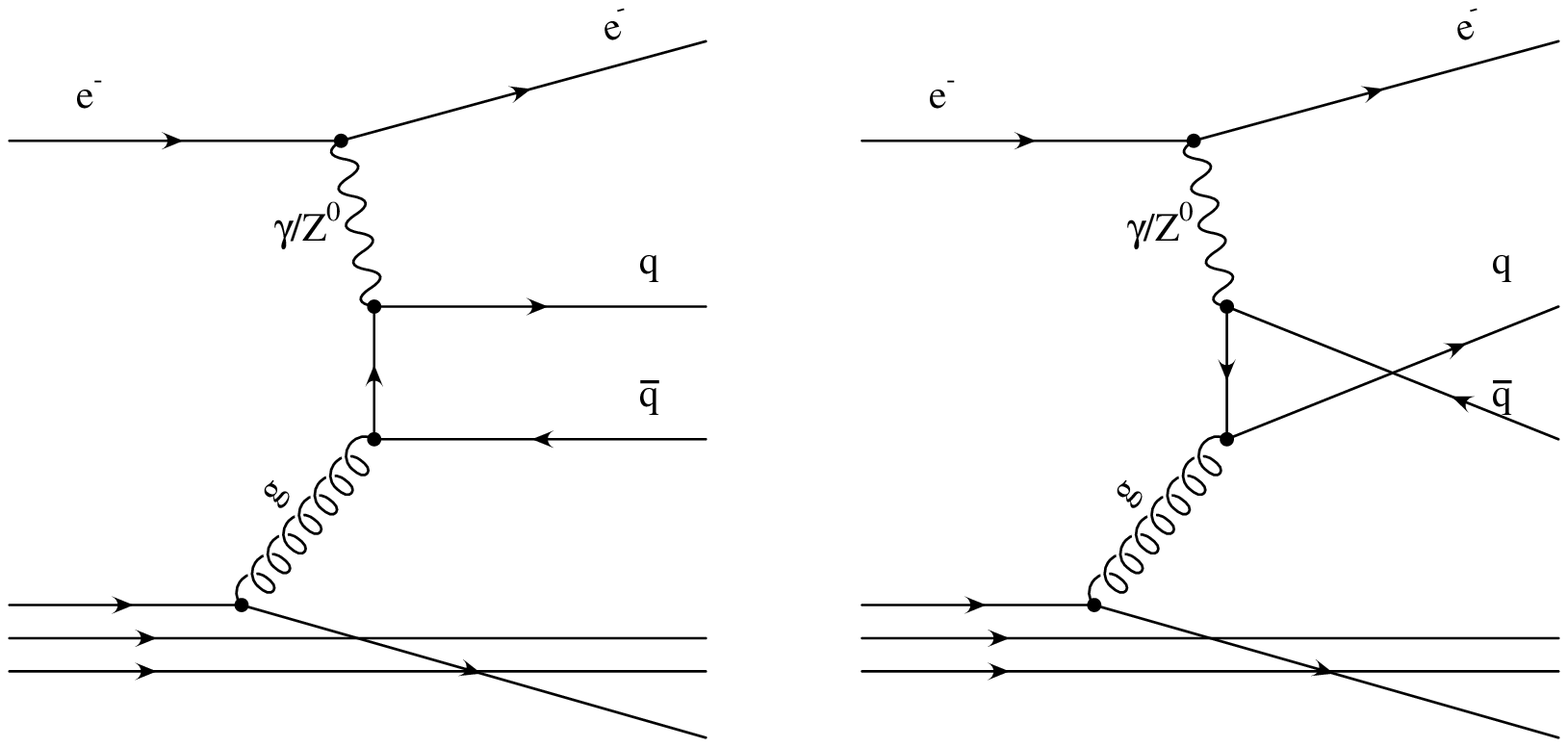,width=4cm,%
    bbllx=68pt,bblly=435pt,bburx=289pt,bbury=661pt,clip=}
   \epsfig{file=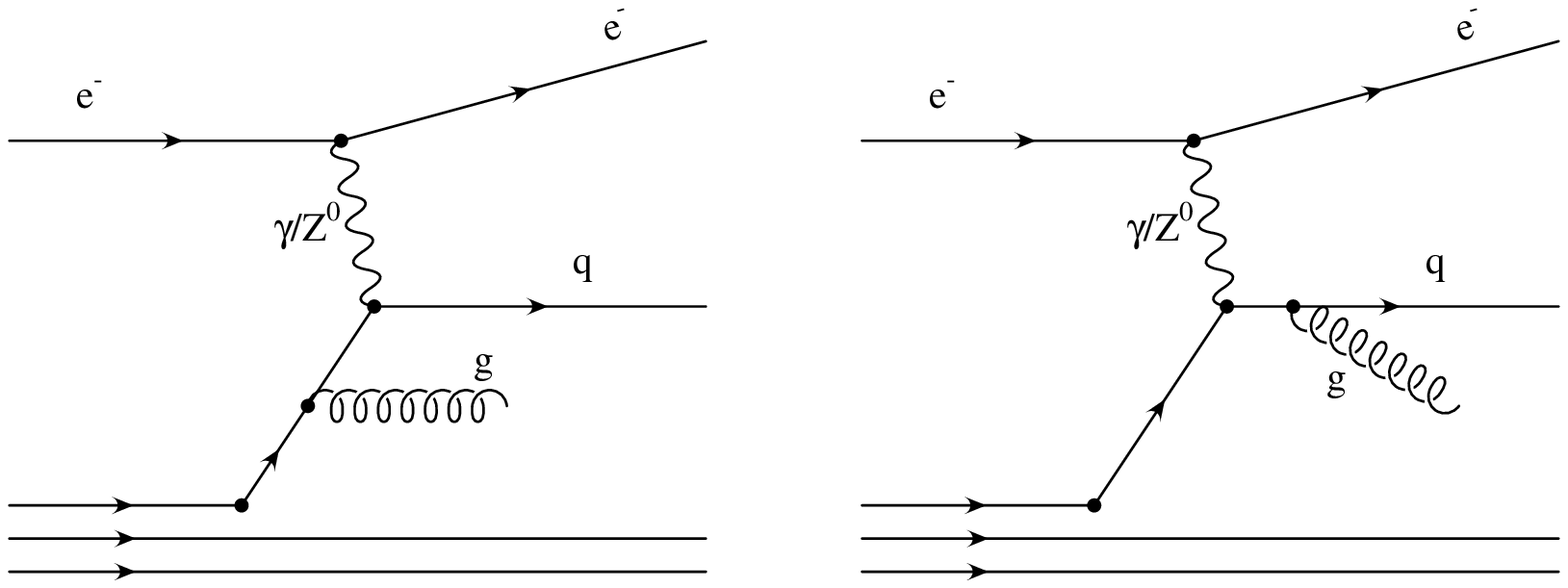,width=4cm,%
    bbllx=60pt,bblly=477pt,bburx=294pt,bbury=661pt,clip=}
}
   \epsfig{file=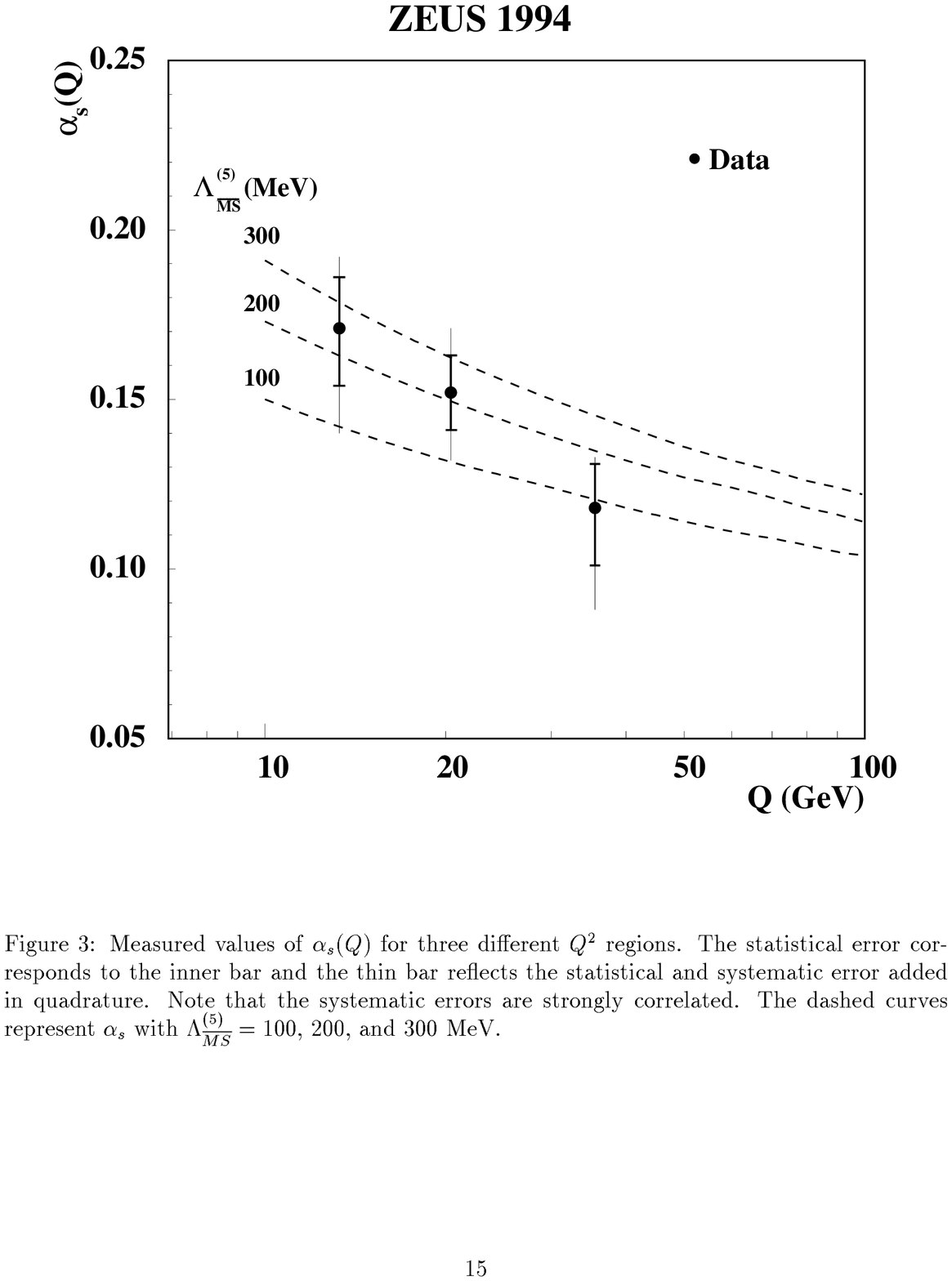,width=6cm,%
   bbllx=87pt,bblly=288pt,bburx=511pt,bbury=716,clip=}
   \scaption{Determination of the strong coupling constant.
             Shown are the boson-gluon fusion (BGF) and the 
             QCD Compton (QCDC) graphs,
             giving rise
             to 2+1 jet events, and the measured
             \as as a function
             of $Q$ from the ZEUS jet rates\cite{z:as}.}
   \label{zeus_as} 
\end{figure}

The processes contributing to jet production in
DIS up to first order in \as are shown
in Figs. \ref{kin},\ref{zeus_as}.
The QPM process results in a so-called
``1+1'' jet topology, while the QCDC and BGF processes give
``2+1'' jet events, where the ``+1'' refers to the unobserved remnant jet.
Since the rate of 2+1 jet production is $\propto \alpha_s$,
$\alpha_s(Q)$ 
can be measured \cite{h1:as,z:as}, and
the scale dependence can be studied.
The ZEUS data of 1994 are shown in (Fig.~\ref{zeus_as}).
The $Q$ dependence is as expected for
a running \as.
A fit yields 
$\alpha_s(m_Z) = 0.117 \pm 0.005 (\rm stat.)
            ^{+0.004}_{-0.005} (\rm syst.)
            \pm 0.007 (\rm theor.)$, 
to be compared with the world average, 
$\alpha_s(m_Z)=0.117\pm 0.005$.
It can be expected that the theoretical uncertainties can
be reduced by studying different jet algorithms \cite{mc:mepjet},
and by obtaining a better understanding of the hadronic final
state dynamics.

\subsection{Thrust in the Breit frame}

In the Breit frame in- and outgoing quark have equal but opposite sign
momenta $Q/2$ (QPM picture),
and in \epem annihilation the outgoing quark and antiquark
have equal but opposite momenta $\sqrt{s}/2=Q/2$.
Due to this similarity it is interesting to compare 
measurements in the Breit current hemisphere in DIS with
\epem data. 
The current hemisphere
of the DIS Breit frame is equivalent to one hemisphere of
the $\epem \rightarrow q\bar{q}$ reaction. 
DIS experiments have the advantage
that they cover a large span in \Qsq
in a single experiment.

\begin{figure}[h]
\vspace{-1cm}
   \centering
 \begin{tabular}{cc}
 \mbox{ \hspace{-1.cm}
 \epsfig{figure=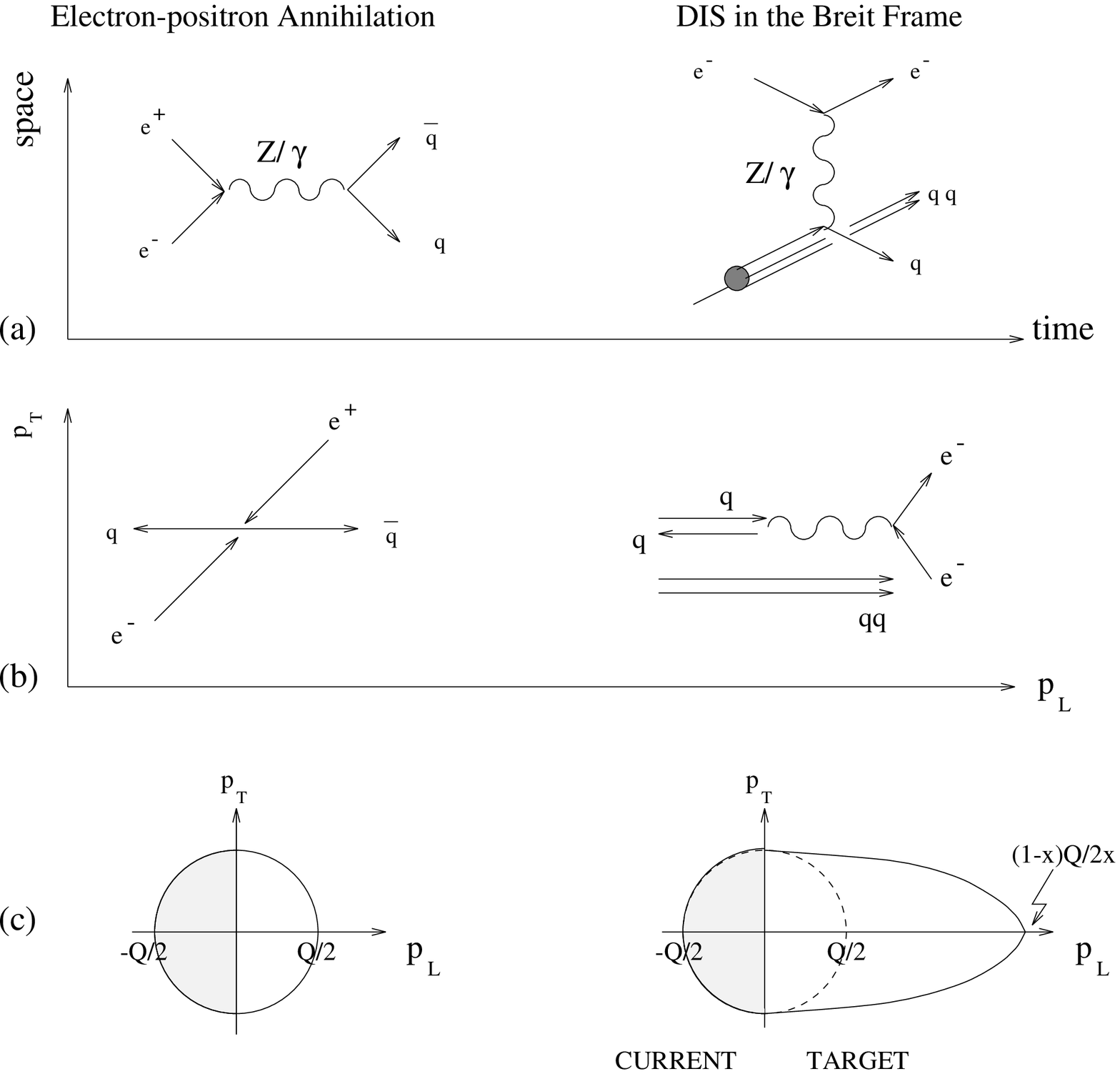,bbllx=211,bblly=300,bburx=335,bbury=513,
         clip=,angle=270,width=4cm}
 \hspace{10cm}
 }
\end{tabular}
\mbox{
   \begin{picture}(1,1) \put(60.,30.){H1 prel.} \end{picture}
   \epsfig{file=tmean.eps,width=7.0cm,%
   bbllx=26pt,bblly=410pt,bburx=468pt,bbury=800,angle=90,clip=}
}
   \hspace{0.5cm}
\mbox{
   \epsfig{file=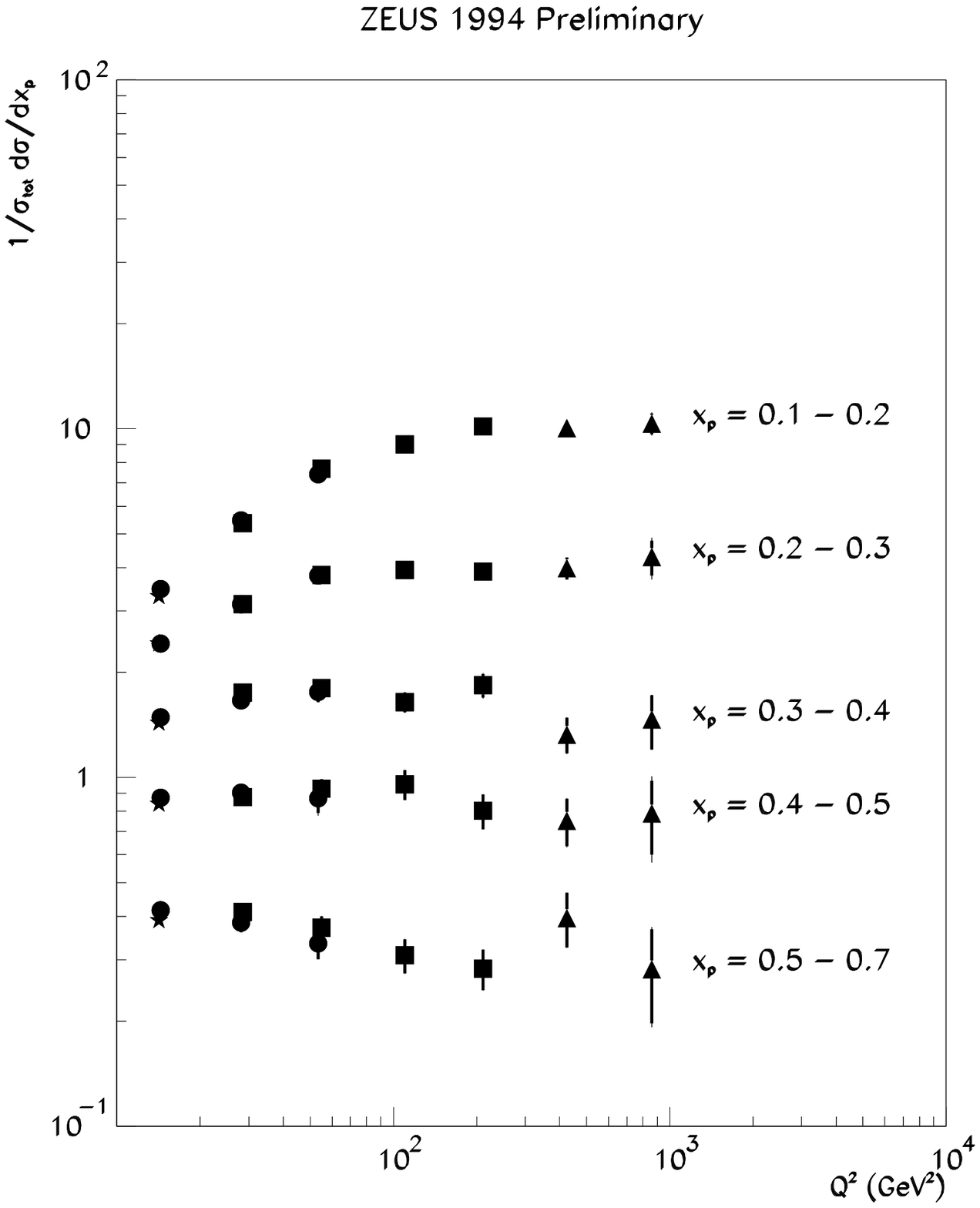,width=7.0cm,%
   bbllx=0pt,bblly=0pt,bburx=450pt,bbury=310}
}   
    \scaption{QCD analysis in the Breit frame (see sketch). 
             {\bf a)} 
             The average $1-T$
             ($T$=thrust) measured in
             the Breit frame as a function of $Q$.
             The data are compared to \epem data and to 
             the LEPTO program, both for hadrons and for partons.
             A QCD fit to the data is also shown. 
             {\bf b)} 
             The normalized charged particle cross section
             in bins of $x_p=2p/Q$ as a function of $Q^2$
             \cite{z:scviol}.}
   \label{breit} 
\end{figure}

Infrared safe
event shape variables have in the past been a very useful way
to measure \as, because it was possible to calculate them in QCD.
One example is the thrust $T$, where $T$ is the normalized sum of all
longitudinal particle momenta 
with respect to the thrust axis $n_T$:
\begin{equation}
T = \max \frac{\sum|\vec{p_i} \cdot \vec{n}_T|}
                 {\sum|\vec{p_i}|}.
\end{equation}
$\vec{n}_T$
is varied to maximize the thrust. $T=1$ when all particles are
collinear, and $T=0.5$ for an isotropic distribution.
The measured thrust value in the Breit current hemisphere 
\cite{h1:thrust}
increases
with increasing $Q$ (i.e. increasing energy of the scattered quark)
-- the current jet becomes more collimated (Fig.~\ref{breit}).
The data agree well with what has been measured in \epem experiments.
The advantage at HERA is evident: the evolution with $Q$ can be
studied in a single experiment. In fact the data can be fit with 
a QCD ansatz 
$\av{1-T} = c_1 \cdot \as(Q) + c_2 \cdot \as^2(Q) + c_3 \cdot 1/Q$,
where the first terms are perturbative up to NLO, and the last term
parametrizes all higher orders and non-perturbative hadronization effects
with a power correction \cite{th:powercorr}. 
The HERA data will allow to check the 
hypothesized universality of the power correction, and to extract
$\alpha_s(Q)$ once the NLO coefficient has been calculated.

\subsection{Scaling violations of charged 
            particle spectra}

The transition from partons to hadrons can be described with
fragmentation functions $D(z,Q)$, which give the probability
to find a hadron carrying the momentum fraction $z$ of the
original parton's momentum. These functions are universal,
and exhibit scaling in lowest order, i.e they are independent
of $Q$. In DIS one can measure the momentum spectra in the 
scaled variable $x_p=p/(Q/2))$, where $p$ is the hadron momentum and
$Q/2$ the maximal possible momentum in the Breit frame.
When QCD radiation is switched on,
the original quark may radiate a gluon, and instead of one hard
parton to fragment there are now two softer ones, which result
in a softer $x_p$ spectrum. These scaling violations increase with
$Q$, as the phase space for QCD radiation increases. 
The preliminary ZEUS data \cite{z:scviol}
on $x_p$  (Fig.~\ref{breit})
indeed show that with increasing $Q$ there are less hadrons
with large $x_p$, and more hadrons with small $x_p$. These
scaling violations are analogous to the scaling violations of
\ftwo, and can be used in a similar fashion to extract \as
\cite{th:scviol}.

\section{Conclusions}

HERA has already delivered a wealth of information about the structure
of the proton, in particular about partons which carry a small
fraction of the proton's momentum. Standard QCD evolution gives
a good description of the inclusive structure function measurements.
Novel QCD effects 
are being searched for in the hadronic final state with some
interesting signals emerging, the interpretation of which is under
discussion. In addition, the strong coupling constant is being
determined using a variety of hadronic final state observables.
These analysis profit from the fact that the scale at which
the coupling constant is being determined
can be varied in a single experiment at HERA.
  
\section{Acknowledgements}

It is a pleasure to thank A.A. Logunov for the kind invitation
to this workshop, and A.V. Kisselev and V.A. Petrov for
their hospitality during the conference. I would also like
to thank my colleagues from H1 and ZEUS for providing me with
the latest data, and J. Dainton, R. Eichler  and R. Klanner for their
careful reading of the manuscript.

\begin{footnotesize}
%
%

\end{footnotesize}
\end{document}